\documentclass[english,aps,prl,amsmath,amssymb,twocolumn,letterpaper,citeautoscript,superscriptaddress,longbibliography]{revtex4-1}

\usepackage{mathtools} %floor and ceiling functions

\usepackage[normalem]{ulem} 
\usepackage{svg}
\usepackage{float}
\usepackage{comment}
\usepackage{amsmath}
\usepackage{physics}
\usepackage{graphicx}
\usepackage{subfigure}
\usepackage{babel}
\usepackage{amstext}
\usepackage{esint}
\usepackage{cases}

\usepackage{hyperref}
\hypersetup{bookmarks=false,linkcolor=blue,urlcolor=blue,colorlinks,citecolor=blue}

\usepackage{pdfpages}	% Packages needed
\usepackage{pgffor}		% to include supplement

\usepackage{accents}

\usepackage{amssymb}
\makeatletter
\AtBeginDocument{\let\LS@rot\@undefined} %This fixes the issue
\makeatother							 %that pdfpages rotates the pages by 90deg.
\usepackage{xcolor}

\usepackage{amsthm}

\newcommand{\prlsection}[1]{{\it {#1}---}}

\begin{document}

\title{Many-Body Entanglement Properties %
of Finite Interacting Fermionic Hamiltonians}%

\author{Irakli Giorgadze}

\affiliation{Department of Physics and Astronomy, Purdue University, West Lafayette, Indiana 47907, USA}

\author{Grayson Welch}

\affiliation{Department of Physics and Astronomy, Purdue University, West Lafayette, Indiana 47907, USA} 
\affiliation{Department of Physics and Engineering, Taylor University, Upland, Indiana 46989, USA}

\author{Haixuan Huang}

\affiliation{Department of Physics and Astronomy, Purdue University, West Lafayette, Indiana 47907, USA}
\affiliation{Perimeter Institute for Theoretical Physics, Waterloo, Ontario N2L 2Y5, Canada}
\affiliation{Department of Physics and Astronomy, University of Waterloo, Waterloo, Ontario N2L 3G1, Canada}

\author{Elio J. K\"onig}
\affiliation{%
Department of Physics, University of Wisconsin-Madison, Madison, Wisconsin 53706, USA}

\author{Jukka I. V\"ayrynen}

\affiliation{Department of Physics and Astronomy, Purdue University, West Lafayette, Indiana 47907, USA}

\date{\today}
\begin{abstract}

We analyze many-body entanglement in interacting fermionic systems by using the $M$-body reduced density matrix. We demonstrate that if a particle number conserving fermionic Hamiltonian contains  only up to $M$-body interaction terms, then its $N$-particle ground state cannot be maximally $M$-body entangled. As a key step in the proof, we show that the energy expectation value of a maximally $M$-body mixed state is equal to the spectral mean of the Hamiltonian on the corresponding $N$-particle subspace. We further demonstrate that the many-body entanglement structure of a ground state can place quantitative constraint on the interaction strength of its parent Hamiltonian. We illustrate the theorem and its implications in Hubbard and extended SYK models. Going beyond ground states, we analyze entanglement generation under unitary dynamics from Slater-determinant initial states in these models. We determine early-time growth and estimate entanglement saturation times. Finally, we derive explicit symmetry-refined saturation upper bounds for $M$-body entanglement in the presence of an Abelian symmetry. % 

\end{abstract}
\maketitle

\prlsection{Introduction}\label{sec:introduction} Quantum entanglement has become a central object for understanding strongly correlated quantum matter~\cite{AmicoVedral2008, LAFLORENCIE2016, EisertPlenio2010}. It has underpinned a large body of work on entanglement entropies, entanglement spectra and their scaling laws in quantum many-body systems, including the area-law paradigm and its violations in gapless phases and Fermi systems~\cite{Hastings2007, KitaevPreskill2006, LevinWen2006, LiHaldane2008, Fidkowski2010, YaoQi2010, Wolf2006,GioevKlich2006, BarthelSchollwoeck2006, LiHaas2006, Swingle2010}. 
Interacting fermions, in particular, are of central importance in quantum chemistry and condensed matter physics. Fermionic many-body systems are realized across a wide range of complex molecules~\cite{HelgakerOlsen2013} and quantum materials, including fractional quantum Hall systems to superconductors and quantum magnets~\cite{Coleman2015, PrangeGirvin1990, SigristUeda1991}. More recently, ultra cold atom arrays have emerged as highly controllable small-scale platforms for simulations of interacting fermionic models~\cite{BlochZwerger2008, GrossBloch2017, CuadraZoller2023}. In parallel, models with all-to-all random interactions, most prominently Sachdev–Ye–Kitaev (SYK)–type Hamiltonians and their extensions, have provided insights in strongly correlated states~\cite{SachdevYe1993, MaldacenaStanford2016, ChowdhurySachdev2022}.

Quantifying the entanglement entropy of identical fermions requires care, since they  have a trivial entanglement associated with the antisymmetrization of the wave function. 
Indeed, some of the commonly used measures assign a large entanglement entropy even to a free fermion ground state~\cite{Wolf2006, GioevKlich2006, BarthelSchollwoeck2006, LiHaas2006, Swingle2010}.
A promising direction in quantifying many-body correlations in fermionic systems is utilizing the particle number based reduced density matrices~\cite{Coleman1963, GigenaRossignoli2021, CarlenReuvers2016, GiorgadzeVayrynen2025, HaqueSchoutens2007, ZozulyaSchoutens2008, HaqueSchoutens2009}, where the system is partitioned in Fock space, instead of real or momentum space. 
Such reduced density matrices remain pure in free fermion ground states. 
For a given $N$-fermion quantum state $\ket{\Psi}$, the $M$-body density matrix (DM) is defined  as
\begin{equation}\label{eq: M-body DM definition}
    \begin{aligned} 
        [\rho^{(M)}_{\Psi}]_{i_1 \cdots i_M, j_1 \cdots j_M} = \bra{\Psi} c^{\dagger}_{j_1} \cdots c^{\dagger}_{j_M} c_{i_M} \cdots c_{i_1} \ket{\Psi},
    \end{aligned}
\end{equation}
using canonical fermion operators $c_{j}, c_{j}^\dagger$, defined in $D$ single-particle orbitals, $j  \in \{1, \dots, D\}$. 
The associated $M$-body entanglement entropy is given by $S(\rho^{(M)}_{\Psi})=-\text{Tr}(\rho^{(M)} _{\Psi}\ln{\rho^{(M)}_{\Psi}})$. 
Slater determinants, which are eigenstates of non-interacting Hamiltonians, have the lowest possible entanglement and by our definition give $S=0$~\cite{EckertLewenstein2002, CarlenReuvers2016}.

\begin{figure*}[t]
    \centering    \includegraphics[width=2\columnwidth]{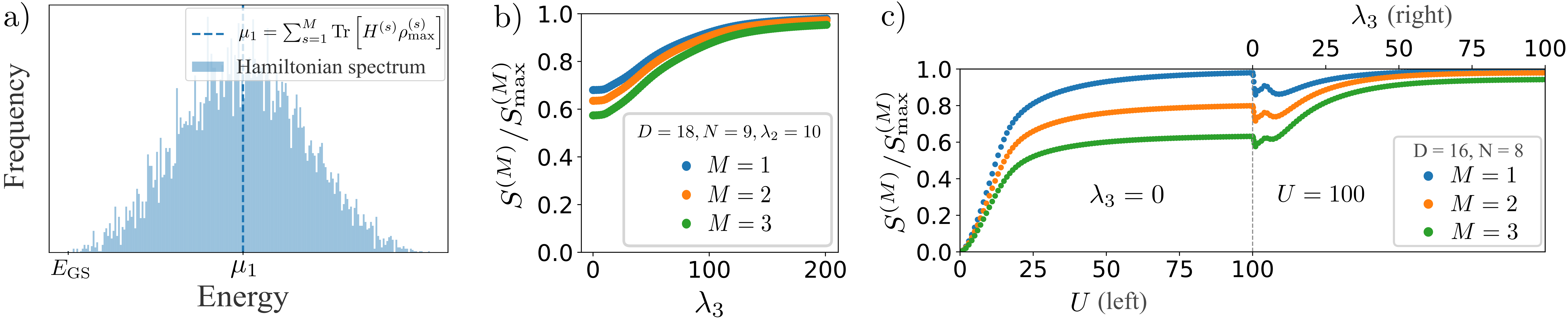}
    \caption{a)~Illustration of the non-existence theorem: When the ground state is maximally $M$-body entangled, the ground state energy $E_{\text{GS}}$ will coincide with the mean $\mu_1$, cf. Eq.~(\ref{eq:MeanEVDist}), implying that the Hamiltonian is trivial. 
    b)~$M$-body entanglement entropy for the ground state of extended complex SYK Hamiltonian with random one, two and three-body interaction terms. We fix $\lambda_1=1, \lambda_2=10$ and change $\lambda_3$, see Eq.~(\ref{eq: SYK extended Hamiltonian}). At $\lambda_3=200$,  $S^{(1)} \approx 0.98S^{(1)}_{\text{max}}$, $S^{(2)} \approx 0.973S^{(2)}_{\text{max}}$ and $S^{(3)} \approx 0.953S^{(3)}_{\text{max}}$. 
    See~\cite{SM} for the same plot with multiple random realizations, which turn out to have a very small spread at the saturation points. 
    c)~$M$-body entanglement entropy of the half-filled triangular lattice Hubbard model [Eq.~(\ref{eq: hubbard model Hamiltonian})] ground state as a function of corresponding coupling parameters. 
    Here, two lattice directions have equal hopping strength $\tau=1$ and the third has $\tau'=2$, and a small magnetic field is applied to break the GS degeneracy. 
    Left sub-panel: Changing two-body interaction $U$, without three-body interaction. 
    Right sub-panel: At $U = 100$, where entropy is saturated but not maximal for $M = 2,3$, we turn on the random three-body interaction term. At $\lambda_3=100$, $S^{(1)} \approx 0.991S^{(1)}_{\text{max}}$, $S^{(2)} \approx 0.979S^{(2)}_{\text{max}}$ and $S^{(3)} \approx 0.942S^{(3)}_{\text{max}}$.  
    }
    \label{fig: combined plots spectral distribution schematic and GS theorem}   
\end{figure*}

Crucially, $M$-body reduced density matrices provide a basis-independent measure of correlations. 
Additionally, they must satisfy the $N$-representability constraints required for compatibility with an underlying many-fermion state~\cite{Coleman1963, Mazziotti2012, Mazziotti2012june, klyachko2004, Li2021}. Historically, most notable is the one-body DM, heavily used in the quantum chemistry setting~\cite{LowdinShull1956, Gilbert1975, MAZZIOTTI2001, LathiotakisMarquess2008} and lately in studies of many-body entanglement in condensed matter~\cite{AikebaierLado2023, VanhalaOjanen2024,ZengXu2002}. More recently, two-body DMs and their cumulants have gained increasing interest as probes of correlations and condensation related phenomena~\cite{Kutzelnigg1999, JuhaszMazziotti2006, SkolnikMazziotti2013, RaeberMazziotti2015, SchoutenMazziotti2022}. They are also connected to experimentally measurable quantities, where they could serve as an entanglement witnesses~\cite{LiuWang2025}.

Vaguely analogous to the volume law of the conventional real-space entanglement entropy, the multi-particle entanglement entropy is saturated when the $M$-body DM of a state $|\Psi\rangle$ is maximally mixed, i.e.~has the form $\rho_{\Psi}^{(M)} = \frac{\text{Tr}\rho_{\Psi}^{(M)}}{n} \mathbb{I}_{n \times n}$, where $n=\binom{D}{M}$ is the size of the DM. In such a case, 
the state can be called \textit{maximally $M$-body entangled}~\cite{GiorgadzeVayrynen2025}. [Note that the above definition of the DM fixes the trace $\text{Tr}\rho^{(M)} = \binom{N}{M}$.]

Some of us recently studied the existence of maximally $M$-body entangled states of $N$ fermions in $D$ orbitals~\cite{GiorgadzeVayrynen2025}. [Such a state does not always exist.] That analysis did not refer to any particular Hamiltonian since the existence is determined by the Hilbert space itself. 
It is then natural to ask how entangled the ground state (GS) of an interacting fermionic Hamiltonian can be. Conversely, to what extent does the entanglement structure of a many-fermion ground state constrain the interactions of the parent Hamiltonian? The first part of this work is concerned with these questions. Surprisingly, we find that the ground states of interacting Hamiltonians cannot exhibit arbitrarily large many-body entanglement. For example, a Hamiltonian with at most two-body interactions cannot have a maximally two-body entangled ground state. We also establish an inequality on the strength of the interaction term inferred from the ground state entanglement structure.

Going beyond ground states, we study the unitary generation of entanglement under time evolution from Slater-determinant initial states. Evolution is carried out under Hubbard and extended SYK models, for which we determine the early-time growth and estimate the entropy saturation times. Finally, we derive symmetry-refined upper bounds for $M$-body entanglement. Here, we make the block structure of $\rho^{(M)}$ explicit and use closed-form expressions for block dimensions and traces to determine the refined bound.

\prlsection{Non-existence of  maximally entangled ground states.} Here we establish the following non-existence theorem: For any finite-dimensional particle number conserving fermionic Hamiltonian with interactions up to order 
$M$ [Eq.~(\ref{eq:M-body interaction Hamiltonian})] which is nontrivial on the $N$-particle subspace, the $N$-particle GS cannot be maximally $M'$-body entangled for any $M' \geq M$.

To see this (details follow), we evaluate the mean of the $N$-particle energy spectra associated to this class of Hamiltonians. We then show that the energy of any maximally $M$-body entangled state equals the spectral mean, Fig.~\ref{fig: combined plots spectral distribution schematic and GS theorem}(a). Thus, unless the Hamiltonian is proportional to the identity in the $N$-particle space, the ground state cannot be maximally $M$-body entangled. Finally, we exploit the fact that any state which is not maximally $M$-body entangled is also not maximally $M'$ body-entangled for $M' \geq M$~\cite{GiorgadzeVayrynen2025}.

We now provide details to the proof outlined above and consider a general finite-dimensional, (up to $M$-body) interacting, and particle number conserving Hamiltonian
\begin{equation}\label{eq:M-body interaction Hamiltonian}
\hat{H}
= \sum_{s=1}^{M}
\sum_{\substack{i_1,\dots,i_s=1 \\ j_1,\dots,j_s=1}}^{D}
H^{(s)}_{i_1 \cdots i_s, j_1 \cdots j_s}
\, c^{\dagger}_{i_1} \cdots c^{\dagger}_{i_s}
c_{j_s} \cdots c_{j_1}.
\end{equation}
We focus on the $N$-particle sector $\mathcal H_N$ of the Fock space where $\hat{H}|_{\mathcal{H}_N}$ is a $\binom{D}{N} \times \binom{D}{N}$ matrix uniquely determined by the set of tensors $H^{(s)}_{i_1 \cdots i_s, j_1 \cdots j_s}$ which are completely anti-symmetric in their first and last $s$ indices. 
The mean of the $N$-particle spectrum, defined as $\mu_1 = \binom{D}{N}^{-1}
\text{Tr}_{\mathcal H_N}[\hat{H}]$,  takes the form~\cite{SM}

\begin{figure*}[t]
    \centering    \includegraphics[width=2\columnwidth]{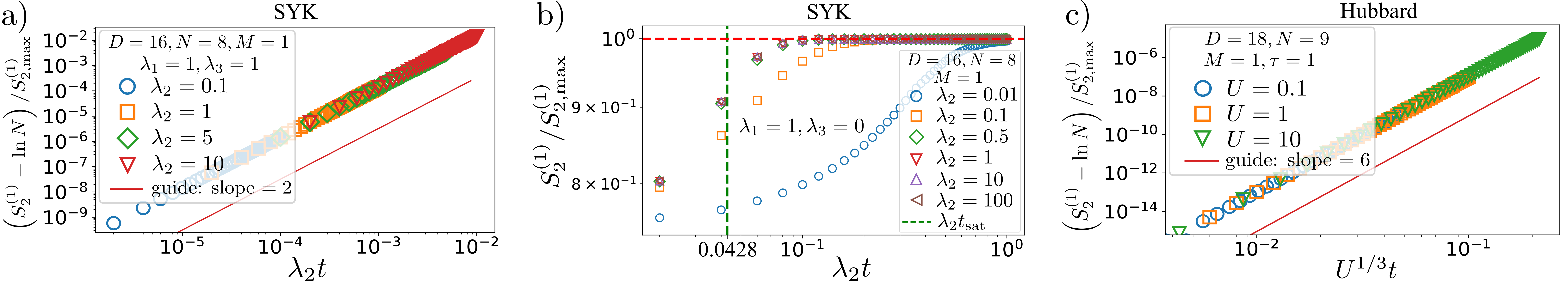}
    \caption{Time evolution of the
    second R\'enyi entropy at half filling for the two-body SYK and  1D Hubbard model dynamics. Initial states are Slater determinants. 
    a)~Early time evolution under two-body SYK Hamiltonian with interaction strength $\lambda_2$, showing $\propto t^2$ scaling for $t\ll \min \{\lambda_2^{-1}, \lambda_1^{-1}\}$.
    b) Entropy growth and saturation for two-body SYK model.
    The vertical dashed green line, obtained from Eq.~(\ref{eq: Entropy saturation time approximation SYK 2-body}) by extrapolating the short-time behavior, lower bounds the saturation time. The saturation time estimate $t_\text{sat}=0.0428/\lambda_2$ is valid when 
    $t_{\text{sat}} \ll \lambda_1^{-1}$. 
    c)~Early time evolution under 1D Hubbard Hamiltonian (hopping $\tau=1$) with on-site interaction $U$, showing $\propto t^6$ scaling  at short times~\cite{SM}.  }
    \label{fig: combined plots earlytime evolution and saturation}   
\end{figure*}

\begin{equation}\label{eq:MeanEVDist}
    \mu_1 = \sum\limits_{s=1}^{M}\frac{\binom{N}{s}}{\binom{D}{s}} \text{Tr}\left(H^{(s)}\right),
\end{equation}
where $H^{(s)}=(s!)^2H^{(s)}_{\rm{flat}}$ and $H^{(s)}_{\rm{flat}}$ denotes the $\binom{D}{s} \times \binom{D}{s}$ matrix obtained by flattening the original antisymmetric tensor $H^{(s)}_{i_1 \cdots i_s, j_1 \cdots j_s}$ in Eq.~(\ref{eq:M-body interaction Hamiltonian}) using composite indices $\alpha = (i_1 < \cdots < i_s)$ and $\beta = (j_1 < \cdots <j_s)$.

We note that each term in the sum of Eq.~(\ref{eq:MeanEVDist}) can be written as $\text{Tr} \left(H^{(s)} \rho^{(s)}_{\text{max}}\right)$, where $\rho^{(s)}_{\text{max}} = \mathbb{I}{\binom{N}{s}}/{\binom{D}{s}} $ and corresponds to the maximally mixed $s$-body reduced DM. 
It is important to note that, depending on the $D, N$ values, there might or might not exist a maximally $M$-body entangled state $\ket{\Psi}$ with $\rho^{(M)}_{\Psi} = \rho^{(M)}_{\rm max}$ in the corresponding Hilbert space $\mathcal H^{(D,N)}$ ~\cite{GiorgadzeVayrynen2025}. But if there exists such a state $\ket{\Psi}$, then $\rho^{(s)}_{\Psi}  = \rho^{(s)}_{\rm max}$ for all $s \leq M$~\cite{GiorgadzeVayrynen2025}. Taking the expectation value $\bra{\Psi}\hat{H} \ket{\Psi}$ of the second quantized Hamiltonian in Eq.~(\ref{eq:M-body interaction Hamiltonian}) then yields the spectral mean of $\hat{H}|_{\mathcal{H}_N}$. Thus, $\ket{\Psi}$ is not the $N$-particle ground state, unless $\hat H$ is trivial in $\mathcal H_N$~\footnote{By means of reflection $\hat H \rightarrow - \hat H$, we note that the same argument applies to the eigenstate corresponding to the maximum $N$-particle eigenvalue of $\hat{H}$.}.

Additionally, the many-body entanglement of a ground state % 
carries information about the
underlying Hamiltonian. To see this, we first note that the difference between the GS energy $E_{\rm GS}$ and the spectral mean $\mu_1$ can be written in terms of reduced DMs as $\mu_1-E_{\rm GS} =
\sum_{s=1}^{M}\text{Tr}\left[
H^{(s)}\Big(\rho^{(s)}_{\max}-\rho^{(s)}_{\rm GS}\Big)
\right]$. Now, the right-hand-side can be bounded using H\"older inequality $\big|\text{Tr}(AB)\big| \le\|A\|_{p}\|B\|_{q}$, where $1/p+1/q=1$. [For an $n \times n$ complex matrix $A$, we denote its $p$-norm as $\Vert A \Vert_p = \big[\text{Tr}(A^\dagger A)^{p/2}\big]^{\frac{1}{p}}$ for $p\in[1, \infty)$.] 
In practice, for a generic Hamiltonian, different 
$s$-body contributions can partially cancel in the sum, so the bound is most informative when one interaction order dominates (or when one can separately estimate the other terms).
In the simple case where the Hamiltonian contains only  $s_0$-body terms, $H^{(s)} = \delta_{s,s_0} H^{(s_0)}$ for $s_0\leq M$, this yields a lower bound on its strength
\begin{equation}\label{eq:Lower bound on interaction}
    \big\|H^{(s_0)}\big\|_{p} \ge
    \frac{(\mu_1 - E_{\rm GS})}{\big\|\rho^{(s_0)}_{\max}-\rho^{(s_0)}_{\rm GS}\big\|_{q}}.
\end{equation}
Put differently, large many-body entanglement of a GS implies a small denominator and thus indicates that the underlying Hamiltonian contains strong high-order interactions.

\prlsection{Examples for the non-existence theorem.} We now illustrate the non-existence theorem for widely studied concrete interacting fermionic Hamiltonians. 

We first consider the extended complex SYK model with up to three-body interaction terms
    \begin{flalign}
         \hat{H}_\text{SYK} \! & = \! \lambda_1 \sum\limits_{i_1j_1}H^{(1)}_{i_1j_1}c^{\dagger}_{i_1}c_{j_1} \!+\! \lambda_2 \! \sum\limits_{i_1i_2j_1j_2} \! H^{(2)}_{i_1i_2,j_1j_2}c^{\dagger}_{i_1}c^{\dagger}_{i_2}c_{j_2}c_{j_1} \nonumber \\ 
        &+ \lambda_3 \sum_{\substack{i_1i_2i_3\\j_1j_2j_3}}H^{(3)}_{i_1i_2i_3,j_1j_2j_3}c^{\dagger}_{i_1}c^{\dagger}_{i_2}c^{\dagger}_{i_3}c_{j_3}c_{j_2}c_{j_1},\label{eq: SYK extended Hamiltonian}
    \end{flalign}
where $H^{(1)}, H^{(2)}, H^{(3)}$ are random Hermitian tensors drawn from a Gaussian distribution with zero mean and variance one.
The parameters $\lambda_1, \lambda_2, \lambda_3$ control the typical magnitude (standard deviation) of the random couplings and therefore tune the relative importance of the different interaction terms. This model serves as an all-to-all, strongly mixing case and one might expect its GS to come closest to maximally three-body entangled structure. Fig.~\ref{fig: combined plots spectral distribution schematic and GS theorem}(b) shows that even though GS is indeed close, it still does not become maximally three-body entangled, consistent with the theorem. 

Next, we consider the Hubbard model with a general Hermitian hopping matrix $\tau_{ij}$
\begin{equation}\label{eq: hubbard model Hamiltonian}
   \hat H_{\rm Hub} = - \sum_{i \neq j; \sigma}\tau_{ij} c^\dagger_{i, \sigma} c_{j, \sigma} + U \sum_{i} \hat n_{i, \uparrow}\hat n_{i, \downarrow}.
\end{equation}
Using Eq.~(\ref{eq:MeanEVDist}), the spectral mean of the Hubbard Hamiltonian on the $N$-particle subspace is $\mu_1 = \frac{\binom{N}{2}}{\binom{D}{2}} \frac{1}{2} D U$, where $U$ is the on-site interaction energy contributing to diagonal elements of $H^{(2)}$, and $D/2$ corresponds to the number of lattice sites. Hopping only contributes to the off-diagonal terms of $H^{(1)}$ and results in $\text{Tr}H^{(1)}$ being zero~\footnote{
Physically, we work in the canonical ensemble and diagonalize $\hat{H}_{\rm Hub}$ in a fixed-$N$ particle sector. Consequently, a chemical potential term $-\mu\hat{N}$ would only shift energies by a constant $-\mu N$ in that sector and hence shift $\mu_1$ by the same constant. At the same time, it would not affect the eigenstates or their entanglement properties discussed in the paper.}.

In the left part of Fig.~\ref{fig: combined plots spectral distribution schematic and GS theorem}(c) we simulate the Hubbard model on a triangular lattice at half-filling with nearest-neighbor hopping. By varying the interaction strength $U$ and computing the GS $M$-body entanglement entropy, we observe that the GS is never maximally two-body entangled for any $U$, in agreement with the theorem. In the right part of Fig.~\ref{fig: combined plots spectral distribution schematic and GS theorem}(c), we fix $U$ and turn on the random three-body interaction controlled by the parameter $\lambda_3$. Now, the GS entanglement entropy saturates at considerably higher values than in the case of Hubbard model alone, but again never becomes  maximally three-body entangled.

Finally, we mention in passing the $t-V$ model studied in Ref.~\cite{HaqueSchoutens2009} which describes the spinless fermions with nearest-neighbor hopping and density-density interactions. As shown there, increasing the interaction strength $V$ increases correlations, yet the two-body entanglement entropy remains far from its maximal value, consistent with our theorem. 

The constraints revealed by the non-existence theorem are limited to ground states (and maximum-energy eigenstates). In order to consider generic states in the Hilbert space, we employ time evolution to generate them and study how $M$-body entanglement behaves in this case.

\prlsection{Time evolution of $M$-body entanglement.} The time evolved $M$-body DM $\rho_{\Psi(t)}^{(M)}$, cf. Eq.~\eqref{eq: M-body DM definition}, is defined through the time-evolution $\ket{\Psi(t)} = e^{- i \hat H t} \ket{\Psi(0)}$ for given initial state $\ket{\Psi(0)}$ and Hamiltonian $\hat H$. The time-evolution of entanglement measure based on $M$-body DM follows accordingly. We start by looking at early times.

\begin{figure}[t]
    \centering    \includegraphics[width=1\columnwidth]{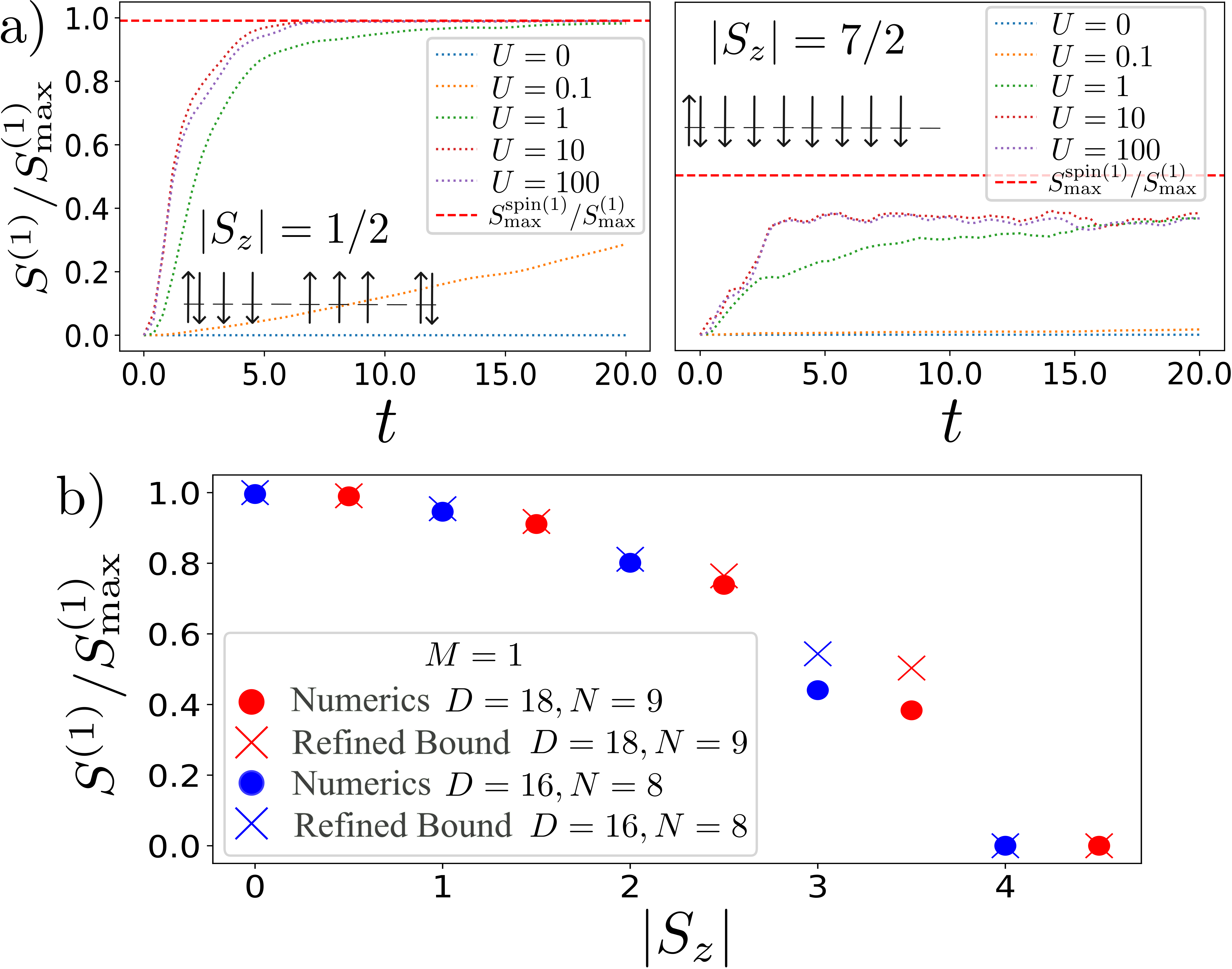}
    \caption{Hubbard dynamics and entropy saturation constrained by symmetries. a)~Time evolution of the one-body entanglement entropy for a 9-site 1D lattice at half-filling, starting from SD initial states in the spin-projection sectors $|S_z|$ of 1/2 and 7/2 (in units of $\hbar$). Symmetry refined upper bounds to the entanglement entropy of Eq.~(\ref{eq: Refined upper bound to entropy}) are given by red dashed lines. 
    Plots for other initial SD states and different $M$ values are provided in~\cite{SM}. 
    b)~Numerically obtained entropy saturation values (circles), similar to panels a), and  theoretical refined bounds (crosses) obtained from Eq.~(\ref{eq: Refined upper bound to entropy}). }
    \label{fig: combined plots refined upper bound}   
\end{figure}

Considering the early time evolution of the entanglement measures, it is convenient to look at the $M$-body DM in the interaction picture and the time dependence of the $2^{\text{nd}}$ R\'enyi entropy $S^{(M)}_{2}(t) = -\ln\text{Tr}[\varrho^{2}(t)]$.
Here, $\varrho(t) = \rho^{(M)}(t)/\binom{N}{M}$ is the trace-normalized $M$-body DM. Note that since the inequality $S^{(M)}_1(\varrho) \geq S^{(M)}_2(\varrho)$ holds in general for trace-normalized DMs, the $2^{\text{nd}}$ R\'enyi entropy lower bounds the von Neumann entropy $S^{(M)}_1$. 

We now present results on the early time evolution for the two models introduced above, with details of the derivation provided in~\cite{SM}. First, for the
extended complex SYK model in Eq.~(\ref{eq: SYK extended Hamiltonian}) with up to two-body interaction terms (i.e., $\lambda_3=0$) and an \textit{arbitrary} Slater determinant (SD) initial state~\cite{SM}, we find that at short times the $2^{\text{nd}}$ R\'enyi entropy obeys
\begin{equation}\label{eq: Entropy scaling SYK 2-body}
    \begin{aligned}
        S^{(M)}_2(t) \approx \ln{\binom{N}{M}} + g_2 \lambda_2^2 t^2,\quad t\ll \min \{\lambda_2^{-1}, \lambda_1^{-1}\},
    \end{aligned}
\end{equation}
where $g_2$ is a constant independent of $\lambda_1,\lambda_2$ and $t$ (it can depend on $D,N,M$), Fig.~\ref{fig: combined plots earlytime evolution and saturation}(a). 

This allows us to estimate the time at which the saturation is approximately
reached by extrapolating the early time evolution. The saturation time in the two-body interaction case is
\begin{equation}\label{eq: Entropy saturation time approximation SYK 2-body}
    \begin{aligned}
        \lambda_2t_{\rm{sat}} = \sqrt{\frac{S^{(M)*}_2 - \ln{\binom{N}{M}}}{g_2}},
    \end{aligned}
\end{equation}
where $S^{(M)*}_2$ is the value of the entropy at saturation. The most natural choice for this estimation is the upper bound itself, i.e $S^{(M)*}_2 = \ln{\binom{D}{M}}$. To obtain the explicit value of the saturation time, we need to find $g_2$, which can be obtained numerically from Fig.~\ref{fig: combined plots earlytime evolution and saturation}(a) with a single data point at an early time. The saturation time Eq.~(\ref{eq: Entropy saturation time approximation SYK 2-body})  is valid when $t_{\text{sat}} \ll \lambda_1^{-1}$ due to the limitation of Eq.~(\ref{eq: Entropy scaling SYK 2-body}). Indeed, Fig.~\ref{fig: combined plots earlytime evolution and saturation}(b) shows good agreement with the above estimate when $\lambda_2 \gtrsim 0.1$,
for $M=1$.
A similar plot for $M=3$ and the discussion for SYK model with three-body interaction term included are provided in Ref.~\cite{SM}. 

More generally, early time expansion~\cite{SM} shows that, starting from an \textit{arbitrary} Slater determinant, the entanglement cannot grow linearly in $t$ and must begin  at least as $t^2$. However, the initial growth can be slower than $t^2$, as illustrated by the Hubbard model. For the Hubbard dynamics, when the initial Slater determinant in the original basis has the form $\ket{\text{SD}} = c^{\dagger}_{i_1\sigma_1}\cdots c^{\dagger}_{i_N\sigma_N} \ket{0}$, we observe $t^6$ behavior, see Fig.~\ref{fig: combined plots earlytime evolution and saturation}(c), while other Slater determinants can show $t^4$ growth. In both cases the dynamics rapidly evolves into the $t^2$ regime.

Finally, Fig.~\ref{fig: combined plots refined upper bound}(a) shows the full time evolution of the entanglement entropy for 1D Hubbard model for two initial SD states. We note that the late-time entropy saturation values differ in the two cases. As shown in Fig.~\ref{fig: combined plots refined upper bound}(b), the entropy saturation values depend on the $z$-component of the spin of the initial SD states.  
In order to understand why different spin sectors saturate to different entanglement plateaus, we now turn to symmetry-refined upper bounds of the entropy.

\prlsection{Symmetry-refined entropy bounds} Our numerical simulations of the entanglement dynamics illustrate that the entanglement entropy saturates at sub-maximal values for models with additional symmetries. 
It is well known that in the presence of  conserved quantities the reduced DM obtained from spatial bipartitions commutes with the corresponding charges, and hence it decomposes into charge sectors~\cite{AresCalabrese2023, LaflorencieRachel2014, GoldsteinSela2018, XavierSierra2018, BonsignoriCalabrese2019, ParezCalabrese2021}. However, the extension to $M$-body DM considered in this work is not immediate. Here we formulate the analogous statement for $M$-body DM and derive its consequences. 

In particular, we concentrate on many-body eigenstates, $\hat{N}_{\sigma} \ket{\Psi} = N_{\sigma} \ket{\Psi}$, of $l$ mutually commuting number operators, $\hat{N}_{\sigma}= \sum_{a = 1}^{D/l} c^{\dagger}_{a\sigma} c_{a\sigma}$ ($\hat{N} = \sum\limits_{\sigma=1}^{l} \hat{N}_{\sigma}$). Here, we decomposed the indices entering the DM in Eq.~\eqref{eq: M-body DM definition} as $i = (a,\sigma)$, with ``site" index $a\in \{1,\dots, D/l\}$ and generalized ``spin'' index $ \sigma \in \{1,\dots, l\}$. In such Abelian symmetry case, we show explicit expressions for the block traces and dimensions of $\rho^{(M)}$, which yield symmetry-refined upper bounds to the entanglement~\cite{SM}.

In the case of spin-1/2 ($l=2$), relevant for the Hubbard model, 
the refined upper bound to the entanglement entropy is given by~\cite{SM} 
\begin{equation}\label{eq: Refined upper bound to entropy}
    \begin{aligned}
        &S^{\text{spin}(M)}_{\text{max}}(N_{\downarrow}, N_{\uparrow}) \\
        &\!= \!\!\!\!\sum\limits_{m_{\downarrow}=\max\{0, M-N_{\uparrow}\}}^{\min \{M, N_{\downarrow}\}} \!\!\binom{N_{\downarrow}}{m_{\downarrow}} \binom{N_{\uparrow}}{M-m_{\downarrow}} \ln \frac{\binom{D/2}{m_{\downarrow}} \binom{D/2}{M-m_{\downarrow}}}{\binom{N_{\downarrow}}{m_{\downarrow}} \binom{N_{\uparrow}}{M-m_{\downarrow}}} .
    \end{aligned}
\end{equation}

Returning to Fig.~\ref{fig: combined plots refined upper bound}(a), the Slater determinant has well-defined $N_{\uparrow}$ and $N_{\downarrow}$ (equivalently $S_z$) and the Hubbard Hamiltonian conserves these quantum numbers. Hence, the refined upper bound on the entropy applies to this evolution, Fig.~\ref{fig: combined plots refined upper bound}(b). Nevertheless, we observe that for larger $S_z$ the entropy does not reach this bound. A likely reason is that the dynamics is further constrained by additional spin structure. In particular, the state has support only in a restricted set of total-spin sectors (with $S \geq |S_z|$), which would lead to a tighter, $S$ refined bound that we do not consider in Eq.~(\ref{eq: Refined upper bound to entropy}).

\prlsection{Conclusions}\label{sec:conclusions} %
In much of quantum physics, one truncates Hamiltonians at quartic interactions, guided by the idea that quadratic plus quartic terms are enough to  generate higher-order interactions, at least perturbatively. Our work shows a limitation of that intuition at the level of ground states of finite-size particle number conserving fermionic Hamiltonians: if the %
Hamiltonian contains  only up to  $M$-particle interactions, then its $N$-particle ground state cannot be maximally $M$-body entangled. 
Even if high-order interactions can appear as effective terms, they cannot make a ground state maximally many-body entangled at the corresponding order. 

Beyond the non-existence theorem, our results suggest a complementary perspective. The many-body entanglement structure of a ground state can place quantitative constraints on the interaction strength of its parent Hamiltonian, Eq.~(\ref{eq:Lower bound on interaction}). In particular, large $M$-body entanglement may require sufficiently strong high-order couplings. 
This may guide the engineering of interacting Hamiltonians that have highly entangled ground states. 

\acknowledgements

\prlsection{Acknowledgements}\label{sec:acknowledgements}
It is a pleasure to thank Sam Garratt, Yao Wang, for useful discussions.
G.W. was supported by the National Science Foundation through the Research Experience for Undergraduates program under Grant Number PHY-2244297. 
Research at Perimeter Institute is supported in part by the Government of Canada through the Department of Innovation, Science and Economic Development and by the Province of Ontario through the Ministry of Colleges,  Universities, Research Excellence and Security. 
Support for this research was provided by the Office
of the Vice Chancellor for Research and Graduate Education at the University of Wisconsin–Madison with
funding from the Wisconsin Alumni Research Founda-
tion (E.J.K.).
E.J.K. thanks the hospitality of the Quantum Science Theory Visitor Program at Purdue University.

\bibliography{refs}

\clearpage
\newpage

%%%%%%%%%% Prefix a "S" to all equations, figures, tables and reset the counter %%%%%%%%%%
\setcounter{equation}{0}
\setcounter{figure}{0}
\setcounter{section}{0}
\setcounter{table}{0}
\setcounter{page}{1}
\makeatletter
\renewcommand{\theequation}{S\arabic{equation}}
\renewcommand{\thepage}{S-\arabic{page}}
\renewcommand{\thesection}{S\arabic{section}}
\renewcommand{\thefigure}{S\arabic{figure}}

% Subfigure references shown as S1(a) when refered it
\renewcommand{\thesubfigure}{(\alph{subfigure})}
\renewcommand{\p@subfigure}{\thefigure} % <-- this is the key line

\renewcommand{\thetable}{S-\Roman{table}}

\renewcommand\section{\@startsection{section}{1}{0pt}%
  {1.5ex plus 1ex minus .2ex}%
  {1ex plus .2ex}%
  {\centering\normalfont\bfseries}}
  
%\renewcommand{\bibnumfmt}[1]{[S#1]}
%\renewcommand{\citenumfont}[1]{S#1}
%%%%%%%%%% Prefix a "S" to all equations, figures, tables and reset the counter %%%%%%%%%%

\begin{widetext}
\begin{center}
Supplementary materials on \\
\textbf{``Many-body Entanglement Properties of Finite Interacting Fermionic Hamiltonians''}\\
Irakli Giorgadze$^{1}$, Grayson Welch$^{1,2}$, Haixuan Huang$^{1,3,4}$, Elio J. K\"onig$^{5}$, Jukka I. V{\"a}yrynen$^{1}$, \\ 
$^{1}$ \textit{Department of Physics and Astronomy, Purdue University, West Lafayette, Indiana 47907, USA}\\
$^{2}$ \textit{Department of Physics and Engineering, Taylor University, Upland, Indiana 46989, USA}\\
$^{3}$ \textit{Perimeter Institute for Theoretical Physics, Waterloo, Ontario N2L 2Y5, Canada}\\
$^{4}$ \textit{Department of Physics and Astronomy, University of Waterloo, Ontario, N2L 3G1, Canada}\\
$^{5}$ \textit{Department of Physics, University of Wisconsin-Madison, Madison, Wisconsin 53706, USA}
\end{center}

These supplementary materials contain details about calculations of the mean of spectrum $\mu_1$ in Sec.~\ref{sec:appendix Nonexistence of max entangled GS} and shows the  ensembles of 1-particle entropies for random 3-body interactions in   Sec.~\ref{sec:appendix SYK ensemble}. Details of the entropy time evolution for extended complex SYK and Hubbard model can be found in Sec.~\ref{sec:appendix Early time evolution}. Finally, in Sec.~\ref{app:Symm}, the proof of the trace and the size of a block in the block diagonal form of the $M$-body DM are provided.

\section{Moments of interacting Hamiltonians}\label{sec:appendix Nonexistence of max entangled GS}

\subsection{Definitions}

We define the $n^{\text{th}}$  moment $\mu_n$ of the $N$-particle spectrum of Hamiltonian $\hat{H}$ [Eq.~(\ref{eq:M-body interaction Hamiltonian}) of the main text] as 
\begin{equation}\label{eq:n-th  moment in terms of eigenstates as a trace appendix}
    \begin{aligned}
        \mu_n = \binom{D}{N}^{-1} 
        \text{Tr}_{\mathcal H_N}[\hat{H}^n],      
    \end{aligned}
\end{equation}
where the trace is over the $\binom{D}{N}$-dimensional $N$-particle sector. 
The complete set of $N$-fermion Slater determinants $\{\ket{\text{SD}_k}\}$  provides a basis of $N$-particle sector. Hence, we can express the $n^{\text{th}}$  moment in the SD basis as 
\begin{equation}\label{eq:n-th  moment in terms of SDs as a trace}
    \mu_n = \binom{D}{N}^{-1}\sum\limits_{k =1}^{\binom{D}{N}}\bra{\text{SD}_{k}}\hat{H}^n \ket{\text{SD}_k}.
\end{equation}
Using the explicit form of the SDs we can obtain different  moments. Although, the evaluation of higher-order moments, particularly
in the presence of higher-order interaction terms in the Hamiltonian, becomes increasingly intricate. 

\subsection{The mean of the eigenvalue distribution}

We start by looking at Hamiltonians with two-body interaction term, i.e. $M=2$  in Eq.~(\ref{eq:M-body interaction Hamiltonian}) of the main text (this could include one-body interaction encoded as two-body interaction):
\begin{equation}\label{eq:2-body interaction Hamiltonian}
    \hat{H} = \sum\limits_{ij,kl=1}^{D} H^{(2)}_{ij,kl} c^{\dagger}_i c^{\dagger}_j  c_l c_k  \,.
\end{equation}
To determine the mean of the eigenvalue distribution, we explicitly plug Eq.~(\ref{eq:2-body interaction Hamiltonian}) in Eq.~(\ref{eq:n-th  moment in terms of SDs as a trace}) for $n=1$
\begin{equation}\label{eq:mean of the eigenval distribution 2-body interaction}
    \mu_1 = \binom{D}{N}^{-1}\sum\limits_{\sigma =1}^{\binom{D}{N}}\bra{\text{SD}_{\sigma}} \sum\limits_{ij,kl=1}^{D} H^{(2)}_{ij,kl} c^{\dagger}_i c^{\dagger}_j  c_l c_k \ket{\text{SD}_{\sigma}}.
\end{equation}
We have $N$-particle Slater determinants of the form $|\text{SD}_{\sigma}\rangle = c^{\dagger}_{i_1} \cdots c^{\dagger}_{i_N} \ket{0}$, with $i_1<i_2 < \cdots < i_N$ and $\sigma = (i_1, i_2, \dots, i_N)$. Additionally, we can flatten the original tensor and look only at ordered pairs $\{i<j\}$, $\{k<l\}$, s.t. 
\begin{equation}\label{eq: Collective index form for 2-body H}
    \begin{aligned}
        \sum\limits_{ij,kl=1}^{D} H^{(2)}_{ij,kl} c^{\dagger}_i c^{\dagger}_j  c_l c_k = (2!)^2\sum\limits_{\alpha,\beta=1}^{\binom{D}{2}} \big(H_{\rm{flat}}^{(2)}\big)_{\alpha\beta} C^{(2)\dagger}_{\alpha} C^{(2)}_{\beta},
    \end{aligned}
\end{equation}
where $\alpha = (i, j)$ is a collective index, $C^{(2)\dagger}_{\alpha} = c^{\dagger}_i c^{\dagger}_j$ is a collective 2-fermion creation operator and $H_{\rm{flat}}^{(2)}$ is $\binom{D}{2} \times \binom{D}{2}$ Hermitian matrix.  In order to determine the mean from Eq.~(\ref{eq:mean of the eigenval distribution 2-body interaction}), we note the following:

1) We will get zero contribution to the sum unless $\{ij\} = \{kl\}$, hence $\alpha = \beta$. 

2) For a given $\{ij\}=\{kl\}$ pair, there can be $\binom{D-2}{N-2}$ remaining $\ket{\text{SD}_{\sigma}}$ Slater determinants in the sum over $\sigma$ that contain this particular pair of orbitals. Note that, $\binom{D-2}{N-2} = \frac{\binom{N}{2}}{\binom{D}{2}} \binom{D}{N}$.

Hence, using the above two remarks,  we obtain
\begin{equation}\label{eq:mean of the eigenval distribution of 2-body interaction final}
    \mu_1 = \frac{\binom{N}{2}}{\binom{D}{2}} \sum\limits_{\alpha=1}^{\binom{D}{2}} (2!)^2 \big(H_{\rm{flat}}^{(2)}\big)_{\alpha\alpha} = \frac{\binom{N}{2}}{\binom{D}{2}} \text{Tr}\left((2!)^2H_{\rm{flat}}^{(2)}\right).
\end{equation}

Now we consider generic $s$-body interacting term in the Hamiltonian given in Eq.~(\ref{eq:M-body interaction Hamiltonian}) and repeat the above process. For $n=1$, Eq.~(\ref{eq:n-th  moment in terms of SDs as a trace}) gives
\begin{equation}\label{eq:mean of the eigenval distribution of s-body interaction term}
    \begin{aligned}
        \mu_1^{(s)} &= \binom{D}{N}^{-1}\sum\limits_{\sigma =1}^{\binom{D}{N}}\bra{\text{SD}_{\sigma}}\sum_{\substack{i_1,\dots,i_s=1 \\ j_1,\dots,j_s=1}}^{D} H^{(s)}_{i_1 \dots i_s, j_1 \dots j_s} c^{\dagger}_{i_1} \cdots c^{\dagger}_{i_s}  c_{j_s} \cdots c_{j_1} \ket{\text{SD}_{\sigma}}.
    \end{aligned}
\end{equation}
We can again flatten the original tensor and look only at ordered tuples of the form $\{i_1< \dots < i_s\}$, $\{j_1 < \dots < j_s\}$, s.t. 
\begin{equation}\label{eq: Collective index form for s-body interactin H}
    \begin{aligned}
        \sum_{\substack{i_1,\dots,i_s=1 \\ j_1,\dots,j_s=1}}^{D} H^{(s)}_{i_1 \dots i_s, j_1 \dots j_s} c^{\dagger}_{i_1} \cdots c^{\dagger}_{i_s}  c_{j_s} \cdots c_{j_1} =(s!)^2 \sum\limits_{\alpha, \beta =1}^{\binom{D}{s}} \big(H_{\rm{flat}}^{(s)}\big)_{\alpha \beta} C^{(s)\dagger}_{\alpha} C^{(s)}_{\beta}, 
    \end{aligned}
\end{equation}
where $\alpha = (i_1 \dots i_s)$ is a collective index, $C^{(s)\dagger}_{\alpha} = c^{\dagger}_{i_1} \cdots c^{\dagger}_{i_s}$ is a collective $s$-fermion creation operator and $H_{\rm{flat}}^{(s)}$ is a $\binom{D}{s} \times \binom{D}{s}$ Hermitian matrix. 

Following the same steps as before, we see that only terms with $\{i_1 \dots i_s\} = \{j_1 \dots j_s\}$ will contribute. 
Now, for a given $\{i_1 \dots i_s\}=\{j_1 \dots j_s\}$ tuple, there can be $\binom{D-s}{N-s}$ remaining $\ket{\text{SD}_{\sigma}}$ Slater determinants in the sum over $\sigma$ in Eq.~(\ref{eq:mean of the eigenval distribution of s-body interaction term}) that contain this particular tuple of orbitals. Again, note that $\binom{D-s}{N-s} = \frac{\binom{N}{s}}{\binom{D}{s}} \binom{D}{N}$. Hence, the contribution to the mean value coming from $s$-body interacting term is
\begin{equation}\label{eq:mean of the eigenval distribution of s-body interaction term final}
    \mu_1^{(s)} = \frac{\binom{N}{s}}{\binom{D}{s}} \sum\limits_{\alpha=1}^{\binom{D}{s}} (s!)^2 \big(H_{\rm{flat}}^{(s)}\big)_{\alpha\alpha} = \frac{\binom{N}{s}}{\binom{D}{s}} \text{Tr}\left((s!)^2H_{\rm{flat}}^{(s)}\right).
\end{equation} 

Finally, since the Eq.~(\ref{eq:mean of the eigenval distribution of s-body interaction term final}) holds for any $s$-body interaction term, for the Hamiltonian containing up to $M$-body interaction terms the mean value is given by
\begin{equation}\label{eq:MeanEVDist Appendix}
    \mu_1 = \sum\limits_{s=1}^{M}\frac{\binom{N}{s}}{\binom{D}{s}} \text{Tr}\left((s!)^2H_{\rm{flat}}^{(s)}\right).
\end{equation}
This concludes the derivation of Eq.~\eqref{eq:MeanEVDist} of the main text.

\section{1-body entanglement entropy for an ensemble of three-body interactions}\label{sec:appendix SYK ensemble}

For completeness, in Fig.~\ref{fig:SYK GS saturation ensemble}, we provide the ensemble realization  for the dependence of the entanglement entropy of the SYK and Hubbard model ground states on the random 3-body interaction. A single realization was shown in Figs.~\ref{fig: combined plots spectral distribution schematic and GS theorem}(b)-(c) of the main text. 

\begin{figure}[h]
    \centering
    \subfigure[]{
        \includegraphics[width=0.45\columnwidth]{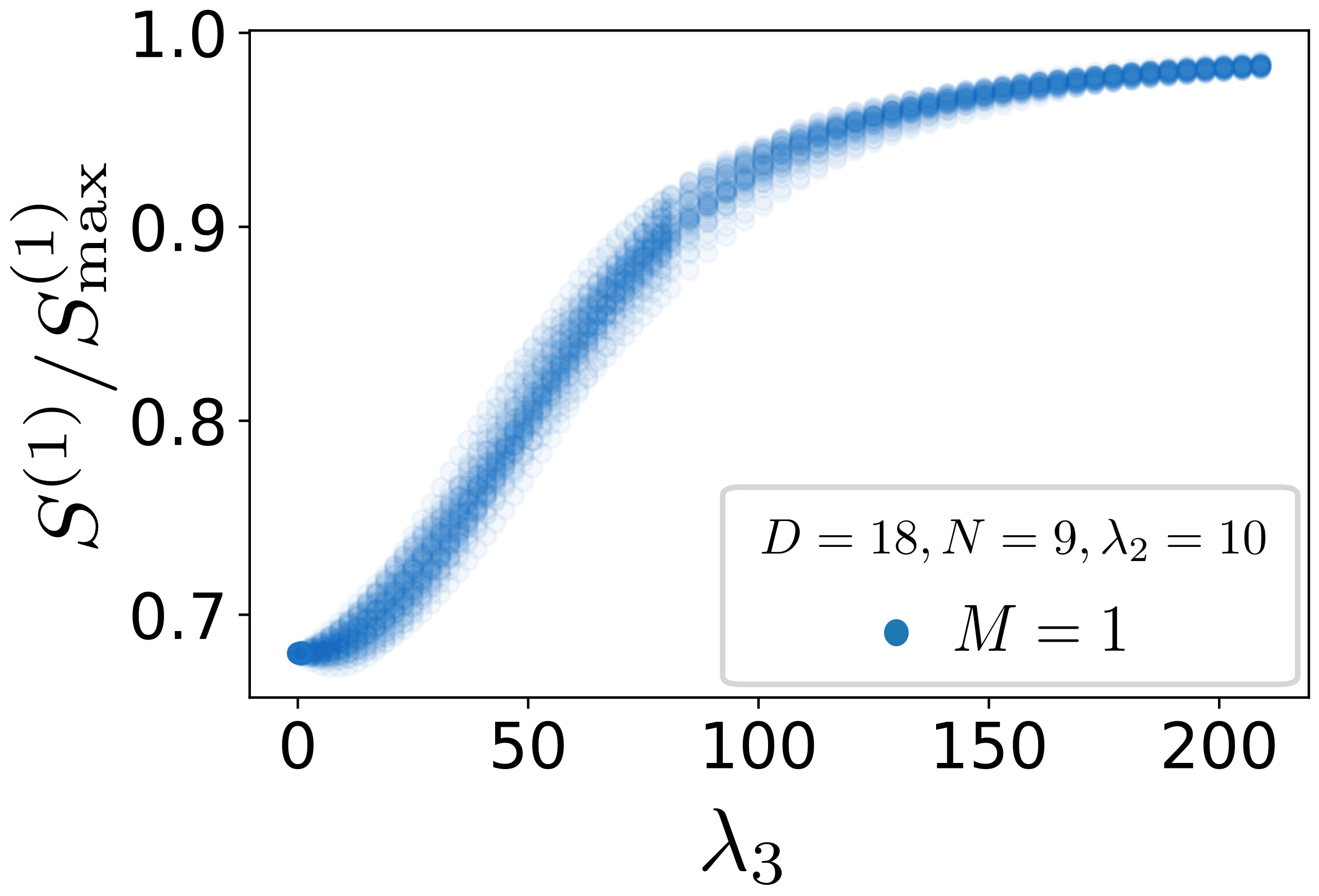}
        \label{fig:SYK only GS saturation ensemble}
    }
    \hspace{-1em}
    \subfigure[]{
        \includegraphics[width=0.45\columnwidth]{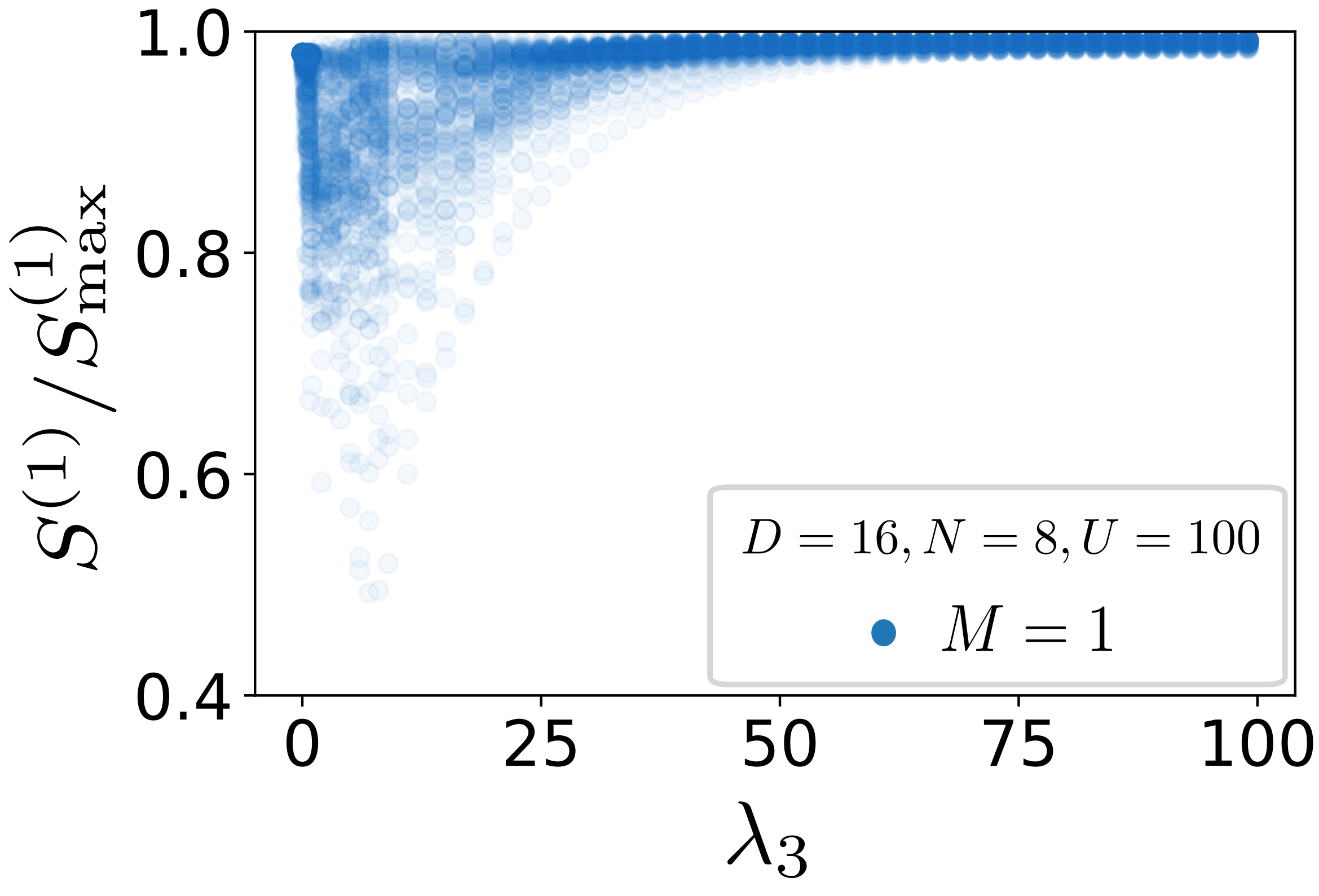}
        \label{fig:Hubbard then SYK GS saturation ensemble}
    }
    \caption{The 1-body entanglement entropy for an ensemble of three-body random terms in the SYK (a) and Hubbard (b) models corresponding to Figs.~\ref{fig: combined plots spectral distribution schematic and GS theorem}(b)-(c) of the main text. 
    The entanglement entropy close to saturation has a very small spread justifying the consideration of a  
     single realization. 
     We used respectively 40 and 100 realizations in (a) and (b). 
    }
    \label{fig:SYK GS saturation ensemble}
\end{figure}

\section{Details of the entropy time evolution}\label{sec:appendix Early time evolution}

In this section, we discuss details of the entropy at long time, Sec.~\ref{appendix time evolution long}, and short times Sec.~\ref{appendix time evolution short general}. 

\subsection{Entropy saturation limited by Hilbert space size}\label{appendix time evolution long}

\begin{figure*}[h]
    \centering
    \subfigure[]{
        \includegraphics[width=0.45\columnwidth]{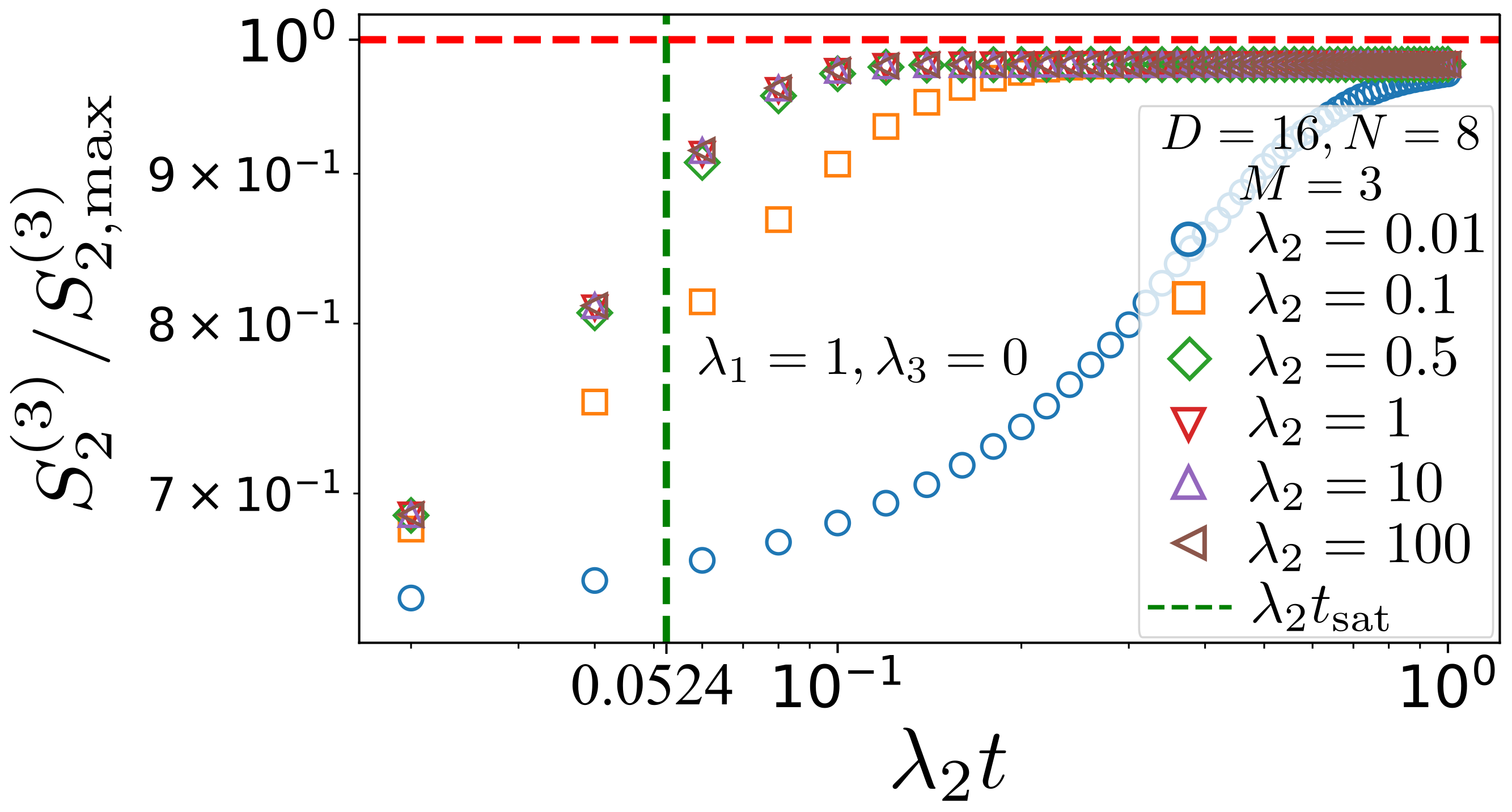}
        \label{fig: Full time evolution 2-body SYK for M=3}
    }
    \subfigure[]{
        \includegraphics[width=0.45\columnwidth]{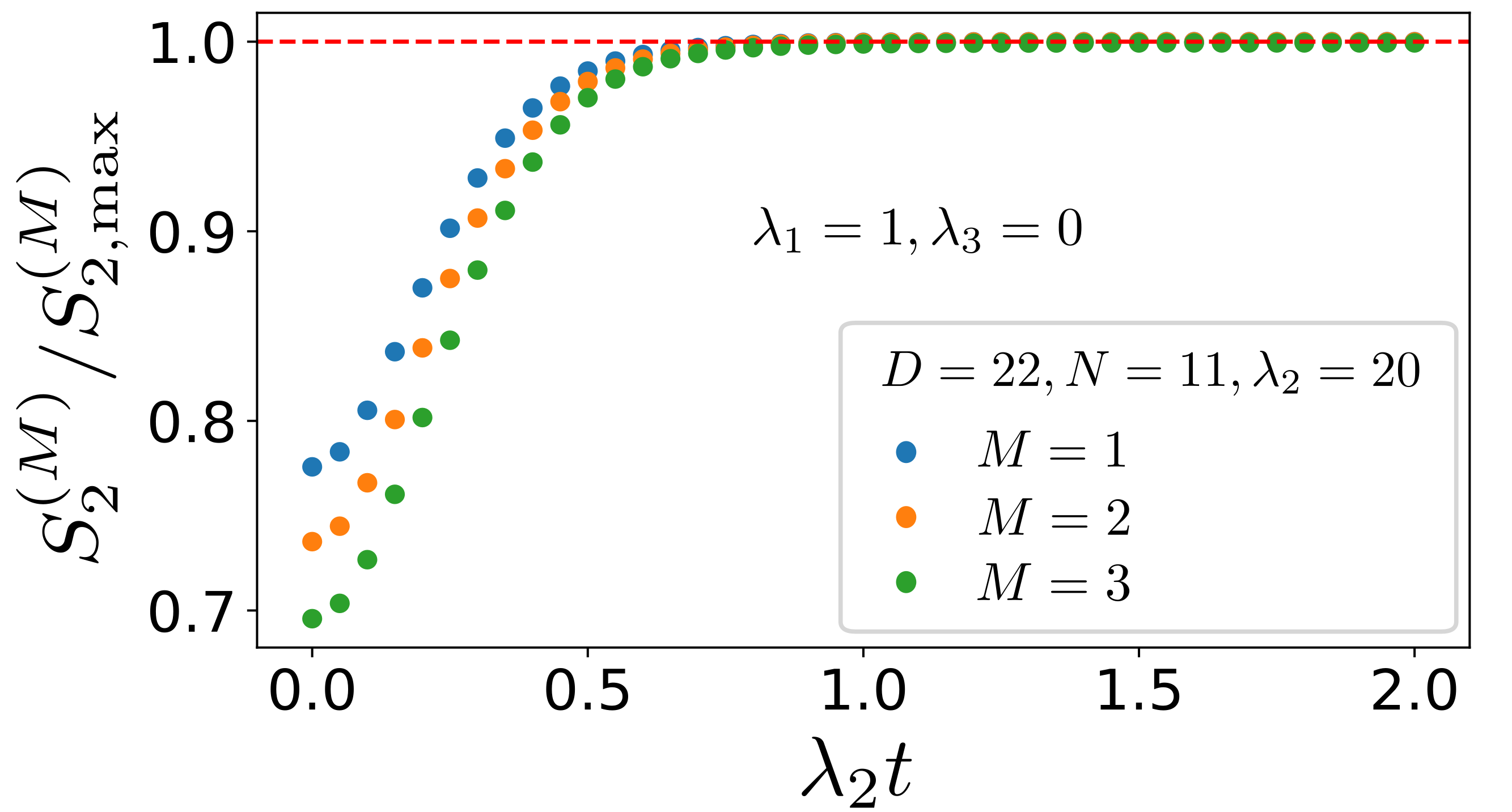}\label{fig: Full time evolution under 2-body SYK for D=22, N=11}
    }
        \subfigure[]{
        \includegraphics[width=0.45\columnwidth]{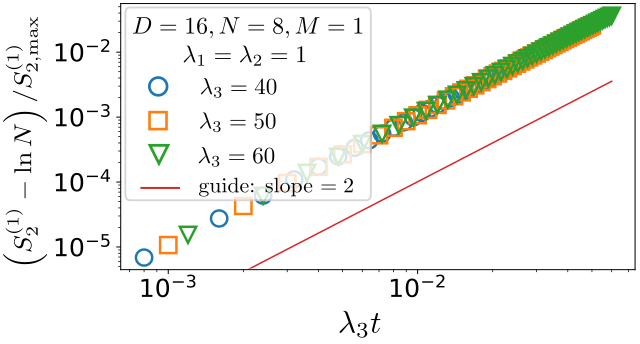}
        \label{fig: Short time evolution 3-body SYK}
    }
    \caption{(a)~Long time evolution of $2^{\text{nd}}$ R\'enyi entropy under two-body SYK dynamics for the case of $M=3$ and $D=16, N=8$. In this case, the upper bound is not saturated. 
    (b)~Long time evolution of $2^{\text{nd}}$ R\'enyi entropy under two-body SYK dynamics for the case $D=22,N=11$. Now, the upper bound is nearly saturated for $M=3$ as well.
    (c)~Early time evolution at half filling of $2^{\text{nd}}$ R\'enyi entropy under three-body SYK dynamics, following a $t^2$ time dependence. The initial state is a Slater determinant (SD). % 
    }
    \label{fig: Early time evolution 3-body SYK}
\end{figure*}

Fig.~\ref{fig: Full time evolution 2-body SYK for M=3} shows the full time evolution of the $2^{\text{nd}}$ R\'enyi entropy for three-body DM and corresponding saturation time estimate. Note that the upper bound on entropy is not saturated under such time evolution. But if we look at larger Hilbert space by increasing the $D, N$ values, we get more states that are close to being maximally $M$-body mixed, as was demonstrated in \cite{GiorgadzeVayrynen2025}. In this case, see Fig.~\ref{fig: Full time evolution under 2-body SYK for D=22, N=11}, the $2^{\text{nd}}$ R\'enyi entropy for three-body DM saturates the upper bound.

\subsection{General considerations about early time evolution}\label{appendix time evolution short general}

The $M$-body DM in the interaction picture takes the form
\begin{equation}\label{eq: M-body DM in interaction picture}
    \begin{aligned}
        \rho^{(M)}_{I_{\alpha \beta}}(t) = {}_{I}\!\bra{\Psi(t)} \left(C^{(M) \dagger}_{\beta} C^{(M)}_{\alpha}\right)_{I}(t) \ket{\Psi(t)}\!_{I},
    \end{aligned}
\end{equation}
where
\begin{equation}\label{eq: State in the interaction picture}
            \ket{\Psi(t)}\!_{I} = e^{i\hat{H}_0t}\ket{\Psi(t)}\!_{S},
    \end{equation}
\begin{equation}\label{eq: State in the Schrodinger picture}
            \ket{\Psi(t)}\!_{S} = e^{-i\hat{H}t}\ket{\Psi(0)},
    \end{equation} 
    and 
\begin{equation}\label{eq: Operators in interaction picture}
            \left(C^{(M) \dagger}_{\beta} C^{(M)}_{\alpha}\right)_{I}(t) = e^{i\hat{H}_0t} C^{(M) \dagger}_{\beta} C^{(M)}_{\alpha} e^{-i\hat{H}_0t},
    \end{equation}
for some initial state $\ket{\Psi(0)}$. From the relation $C^{(M)}_{\alpha} = c_{i_1} \cdots c_{i_M}$ and the fact that $\hat{H}_0$ is a non-interacting part of the full Hamiltonian $\hat{H} = \hat{H}_0 + \hat{V}$, the interaction picture simply gives a single-particle basis transformation of the creation/annihilation operators. This in turn transforms the $M$-body DM unitarily and does not change its eigenvalues or the entropy~\cite{GiorgadzeVayrynen2025, GigenaRossignoli2021}. Hence, for the purposes of calculating the entropy, we can still work in the basis $C^{(M)}_{\alpha}$ to write
\begin{equation}\label{eq: DM with only state in interaction picture}
    \begin{aligned}
        \rho^{(M)}_{\alpha \beta}(t) = {}_{I}\!\bra{\Psi(t)} C^{(M)\dagger}_{\beta} C^{(M)}_{\alpha} \ket{\Psi(t)}\!_{I}.
    \end{aligned}
\end{equation}
and consider the $2^{\text{nd}}$ R\'enyi entropy 
\begin{equation}\label{eq: 2nd Renyi entropy}
    \begin{aligned}
        S^{(M)}_{2}(t) = -\ln{\text{Tr}[\rho^{(M)}(t)^2/\binom{N}{M}}^2],
    \end{aligned}
\end{equation}
where we divide the $M$-body DM by $\binom{N}{M}$ for trace-normalization. To obtain an early time evolution of the $2^{\text{nd}}$ R\'enyi entropy, we expand it in small time near $t=0$, retaining the first non-zero expansion term
\begin{equation}\label{eq: 2nd Renyi entropy taylor expansion}
    \begin{aligned}
        S^{(M)}_2(t) \approx& \ln{\binom{N}{M}}
        + \binom{N}{M}^{-1} \left[\frac{d^m}{dt^m}\text{Tr}\rho^{(M)}(t)^2|_{t=0} \right] \frac{t^m}{m!},
    \end{aligned}
\end{equation}
with $m=1,2, \dots$ non-zero positive integer.
Therefore, we need to consider the time derivatives of $\text{Tr}\rho^{(M)}(t)^2$. For notational simplicity let us replace $\rho^{(M)}(t) \equiv \rho(t)$ in the following calculations. Looking at the first time derivative and taking Slater determinant (SD) as an initial $N$-fermion state $\ket{\Psi(0)} = \ket{\text{SD}} \equiv c^{\dagger}_{i_1} \cdots c^{\dagger}_{i_N} \ket{0}$, we obtain
\begin{equation}\label{eq: First time derivative of trace}
    \begin{aligned}
        \frac{d}{dt}\text{Tr}\rho(t)^2|_{t=0} = 2 \text{Tr}(\rho \dot{\rho}) |_{t=0} = 0,
    \end{aligned}
\end{equation}
for any Hamiltonian. To see this, first we note that $\rho|_{t=0}$ is a diagonal matrix with $\binom{N}{M}$ 1-s and the remaining values zeros. On the other hand, using the interaction picture Schr\"odinger equation $i\frac{d}{dt}\ket{\Psi(t)}\!_{I} = \hat{V}_{I}(t) \ket{\Psi(t)}\!_{I}$,
\begin{equation}\label{eq: First time derivative of DM rho^M}
    \begin{aligned}
        \dot{\rho}_{\alpha \beta} = \frac{1}{i} {}_{I}\!\langle [C^{(M)\dagger}_{\beta} C^{(M)}_{\alpha}, \hat{V}_I(t)] \rangle\!_{I},
    \end{aligned}
\end{equation}
with the diagonal terms vanishing at $t=0$. From these two facts, Eq.~(\ref{eq: First time derivative of trace}) follows. Note that even though the current argument was for the case when the SD state was in the basis where $\rho|_{t=0}$ is diagonal, the vanishing of first order time derivative is independent of the single particle basis, as long as the initial state is the SD state. This is due to the cyclic property of the trace and the fact that the eigenvalues of the $M$-body DM are still 1-s and 0-s, while the diagonal elements of $\dot{\rho}|_{t=0}$ are still zero in the new basis.

Now we can look at the second time derivative
\begin{equation}\label{eq: Second time derivative of trace}
    \begin{aligned}
        \frac{d^2}{dt^2}\text{Tr}\rho(t)^2|_{t=0} = 2 \text{Tr}(\dot{\rho}^{2} + \rho \ddot{\rho}) |_{t=0},
    \end{aligned}
\end{equation}
with
\begin{equation}\label{eq: Second time derivative of DM rho^M}
    \begin{aligned}
        \ddot{\rho}_{\alpha \beta} = & \frac{1}{i} {}_{I}\!\langle [C^{(M)\dagger}_{\beta} C^{(M)}_{\alpha}, \dot{\hat{V}}_I(t)] \rangle\!_{I}+ \frac{1}{(i)^2} {}_{I}\!\langle [[C^{(M)\dagger}_{\beta} C^{(M)}_{\alpha}, \hat{V}_I(t)], \hat{V}_I(t)] \rangle\!_{I}.
    \end{aligned}
\end{equation}
Eq.~(\ref{eq: Second time derivative of trace}) no longer evaluates to zero in general, even for the SD states, and depends on the Hamiltonian in question. Similar to the $\dot\rho|_{t=0}$, that has zero diagonal elements, the first term in  $\ddot\rho|_{t=0}$ in Eq.~(\ref{eq: Second time derivative of DM rho^M}) vanishes when $\alpha=\beta$. Hence, the first term in Eq.~(\ref{eq: Second time derivative of DM rho^M}) combined with $\rho|_{t=0}$ in Eq.~(\ref{eq: Second time derivative of trace})   evaluates to zero. Only the second term of $\ddot\rho|_{t=0}$ contributes to Eq.~(\ref{eq: Second time derivative of trace}). 
When that term is non-zero, the $2^{\text{nd}}$ R\'enyi entropy increases quadratically in $t$ for short times. 

Next, we consider the SYK and Hubbard models as concrete examples.

\subsection{Extended SYK model early time evolution}

We consider the extended complex SYK model with up to three-body interaction terms
\begin{equation}\label{eq: SYK extended Hamiltonian Appendix}
    \begin{aligned}
        \hat{H}_{\text{SYK}} = & \lambda_1 \sum\limits_{i_1j_1}H^{(1)}_{i_1j_1}c^{\dagger}_{i_1}c_{j_1}
        + \lambda_2 \sum\limits_{i_1i_2j_1j_2}H^{(2)}_{i_1i_2,j_1j_2}c^{\dagger}_{i_1}c^{\dagger}_{i_2}c_{j_2}c_{j_1}
        + \lambda_3 \sum_{\substack{i_1i_2i_3\\j_1j_2j_3}}H^{(3)}_{i_1i_2i_3,j_1j_2j_3}c^{\dagger}_{i_1}c^{\dagger}_{i_2}c^{\dagger}_{i_3}c_{j_3}c_{j_2}c_{j_1},
    \end{aligned}
\end{equation}
where $H^{(1)}, H^{(2)}, H^{(3)}$ are random Hermitian tensors drawn from a Gaussian distribution with zero mean and variance one. In this example the interaction term is
\begin{equation}\label{eq: SYK interaction term only}
    \hat{V} = \lambda_2 \hat{V}_2
        + \lambda_3 \hat{V}_3,
\end{equation}
where
\begin{subequations}
\begin{align}
    \hat{V}_2 &= \sum\limits_{i_1i_2j_1j_2}H^{(2)}_{i_1i_2,j_1j_2}c^{\dagger}_{i_1}c^{\dagger}_{i_2}c_{j_2}c_{j_1},\\
    \hat{V}_3 &=  \sum_{\substack{i_1i_2i_3\\j_1j_2j_3}}H^{(3)}_{i_1i_2i_3,j_1j_2j_3}c^{\dagger}_{i_1}c^{\dagger}_{i_2}c^{\dagger}_{i_3}c_{j_3}c_{j_2}c_{j_1}.\label{eq: SYK interaction term V_3}
\end{align}
\end{subequations}

Eq.~(\ref{eq: Second time derivative of trace}) does not evaluate to zero for SYK model, hence the $2^{\text{nd}}$ R\'enyi entropy evolves as $t^2$ for small times. We are interested in the scaling of the entropy with the $\lambda_2$ and $\lambda_3$ parameters for short times. From Eq.~(\ref{eq: Second time derivative of trace}) and the form of $\rho|_{t=0}$, $\dot\rho|_{t=0}$ in Eq.~(\ref{eq: First time derivative of DM rho^M}) and the second term of $\ddot\rho|_{t=0}$ in Eq.~(\ref{eq: Second time derivative of DM rho^M}), we see that 
\begin{equation}\label{eq: Second time derivative of trace SYK case}
    \begin{aligned}
        \frac{d^2}{dt^2}\text{Tr}\rho(t)^2|_{t=0} = \lambda_2^2 a_1 + \lambda_3^2 a_2 + \lambda_2 \lambda_3 a_3,
    \end{aligned}
\end{equation}
for some constants $a_1, a_2, a_3$ not depending on $\lambda_2,\lambda_3$. One can see this by isolating $\lambda_2, \lambda_3$ dependence in $\dot\rho|_{t=0}$ and second term of $\ddot\rho|_{t=0}$ 
\begin{equation}\label{eq:  First time derivative of DM rho^M in SYK}
    \begin{aligned}
        \frac{1}{i} {}_{I}\!\langle [C^{(M)\dagger}_{\beta} C^{(M)}_{\alpha}, \hat{V}_I(t)] \rangle\!_{I} |_{t=0} = \lambda_2\, {}_{I}\!\langle [C^{(M)\dagger}_{\beta} C^{(M)}_{\alpha}, \hat{V}_{2,I}(t)] \rangle\!_{I} |_{t=0} + \lambda_3 \, {}_{I}\!\langle [C^{(M)\dagger}_{\beta} C^{(M)}_{\alpha}, \hat{V}_{3,I}(t)] \rangle\!_{I} |_{t=0},
    \end{aligned}
\end{equation}
\begin{equation}\label{eq: Second time derivative of DM rho^M in SYK}
    \begin{aligned}
        \frac{1}{(i)^2} {}_{I}\!\langle [[C^{(M)\dagger}_{\beta} C^{(M)}_{\alpha}, \hat{V}_I(t)], \hat{V}_I(t)] \rangle\!_{I}|_{t=0}
        = \lambda_2^{2} \hat{A}_{1_{\alpha \beta}} + \lambda_3^2 \hat{A}_{2_{\alpha\beta}} + \lambda_2\lambda_3\hat{A}_{3_{\alpha \beta}},
    \end{aligned}
\end{equation}
where $\hat{A}_{1_{\alpha \beta}}, \hat{A}_{2_{\alpha \beta}}, \hat{A}_{3_{\alpha \beta}}$ are matrices obtained from nested commutators of $C^{(M)\dagger}_{\beta} C^{(M)}_{\alpha}$ with $\hat{V}_2$ and $\hat{V}_3$.

In particular, looking at only two-body interaction by setting $\lambda_3 = 0$ in Eq.~(\ref{eq: Second time derivative of trace SYK case}), we obtain the entropy scaling with $\lambda_2$ as in Eq.~(\ref{eq: Entropy scaling SYK 2-body}) of the main text.

Similarly, fixing $\lambda_2$ and looking at the limit $\lambda_3 \gg \lambda_2$ in Eq.~(\ref{eq: Second time derivative of DM rho^M in SYK}), for short times the entropy scales with $\lambda_3$ as 
\begin{equation}\label{eq: Entropy scaling SYK 3-body}
    \begin{aligned}
        S^{(M)}_2(t) \approx \ln{\binom{N}{M}} + g_3 \lambda_3^2 t^2, \quad t\ll \min \{\lambda_3^{-1}, \lambda_1^{-1}\}
    \end{aligned}
\end{equation}
where $g_3$ is some constant independent of $\lambda_3$ and $t$, Fig.~\ref{fig: Short time evolution 3-body SYK}.
We can again extrapolate to estimate the time at which the saturation is approximately reached. The saturation time in the three-body interaction case is
\begin{equation}\label{eq: Entropy saturation time approximation SYK 3-body}
    \begin{aligned}
        \lambda_3t_{\rm{sat}} = \sqrt{\frac{S^{(M)*}_2 - \ln{\binom{N}{M}}}{g_3}},
    \end{aligned}
\end{equation}
where $S^{(M)*}_2$ is the value of the entropy at saturation. To obtain the explicit value of the saturation time, we need to find $g_3$, which can be obtained numerically from Fig.~\ref{fig: Short time evolution 3-body SYK} with a single data point at an early time. The saturation time in Eq.~(\ref{eq: Entropy saturation time approximation SYK 3-body}) is valid when $t_{\text{sat}} \ll \min \{\lambda_2^{-1}, \lambda_1^{-1}\}$ due to the limitation of Eq.~(\ref{eq: Entropy scaling SYK 3-body}). 

\subsection{1D Hubbard model  early time evolution}

The 1D Hubbard model Hamiltonian reads
\begin{equation}\label{eq: Hubbard Hamiltonian in 1D}
    \begin{aligned}
        \hat{H}_{\text{Hub}} = -\tau \sum\limits_{i \sigma}(c^{\dagger}_{i,\sigma}c_{i+1,\sigma} + c^{\dagger}_{i+1,\sigma}c_{i,\sigma}) + U\sum\limits_{i}n_{i\uparrow}n_{i\downarrow},
    \end{aligned}
\end{equation}
where $\tau$ is the hopping strength and $U$ is the on-site interaction strength. The interaction term is $\hat{V}=U\sum\limits_{i}n_{i\uparrow}n_{i\downarrow}$ in this example.

Considering the case of one-body DM, we can show that $\dot\rho^{(1)}|_{t=0} = 0$. Indeed,  
\begin{equation}\label{eq: First time derivative of DM rho^1 for Hubbard}
    \begin{aligned}
        \dot{\rho}^{(1)}_{i\sigma,j\sigma'}|_{t=0} &= \frac{1}{i} \bra{\text{SD}} [c^{\dagger}_{j\sigma'} c_{i\sigma}, U\sum\limits_{k}n_{k\uparrow}n_{k\downarrow}] \ket{\text{SD}} = \frac{1}{i} \bra{\text{SD}}c^{\dagger}_{j\sigma'}(n_{i,-\sigma} - n_{j, -\sigma'})c_{i\sigma}\ket{\text{SD}} = 0,
    \end{aligned}
\end{equation}
for all $i\sigma, j\sigma'$ values when $\ket{\text{SD}} = c^{\dagger}_{i_1\sigma_1}\cdots c^{\dagger}_{i_N\sigma_N} \ket{0}$.
Similarly, considering the second term of $\ddot{\rho}^{(1)}|_{t=0}$ we have
\begin{equation}\label{eq: Second time derivative of DM rho^1 for Hubbard second term}
    \begin{aligned}
        &\frac{1}{(i)^2} \bra{\text{SD}} [[c^{\dagger}_{j\sigma'} c_{i\sigma}, \hat{V}], \hat{V}] \ket{\text{SD}}=\bra{\text{SD}}c^{\dagger}_{j\sigma'}(n_{i,-\sigma} - n_{j, -\sigma'})^2c_{i\sigma}\ket{\text{SD}} = 0,
    \end{aligned}
\end{equation}
for all $i\sigma, j\sigma'$ values.
The first term of $\ddot\rho^{(1)}|_{t=0}$ has the form
\begin{equation}\label{eq: Second time derivative of DM rho^1 for Hubbard first term}
    \begin{aligned}
        &\frac{1}{i} {}_{I}\!\langle [c^{\dagger}_{j\sigma'} c_{i\sigma}, \dot{\hat{V}}_I(t)] \rangle\!_{I} |_{t=0} = \frac{1}{i} \bra{\text{SD}} [c^{\dagger}_{j\sigma'} c_{i\sigma}, i[H_0, \hat{V}]]\ket{\text{SD}},
    \end{aligned}
\end{equation}
where
\begin{equation}\label{eq: H_0 V commutator Hubbard}
    \begin{aligned}
        [H_0,\hat{V}] = -\tau U\sum\limits_{k,\rho}(c^{\dagger}_{k,\rho}c_{k+1,\rho}-c^{\dagger}_{k+1,\rho}c_{k,\rho})(n_{k+1,-\rho} - n_{k, -\rho})
    \end{aligned}
\end{equation}
and the diagonal elements are evaluated to zero as before. From these we can see that second order time derivative in Eq.~(\ref{eq: Second time derivative of trace}) evaluates to zero for the Hubbard model and initial SD state in the original basis. Therefore, we look at higher order time derivatives. Combining with the above results, for third order time derivative, we get
\begin{equation}\label{eq: Third time derivative of trace}
    \begin{aligned}
        \frac{d^3}{dt^3}\text{Tr}\rho(t)^2|_{t=0} = 2 \text{Tr}(3\dot{\rho}\ddot{\rho} + \rho \dddot{\rho}) |_{t=0} = 0,
    \end{aligned}
\end{equation}
where
\begin{equation}\label{eq: Third time derivative of DM rho^1 for Hubbard}
    \begin{aligned}
        \dddot{\rho}_{i\sigma, j\sigma'} = & \frac{1}{i} {}_{I}\!\langle [C^{(M)\dagger}_{j\sigma'} C^{(M)}_{i\sigma}, \ddot{\hat{V}}_I(t)] \rangle\!_{I} + \frac{1}{(i)^3} {}_{I}\!\langle [[[C^{(M)\dagger}_{j\sigma'} C^{(M)}_{i\sigma}, \hat{V}_I(t)], \hat{V}_I(t)], \hat{V}_I{t}] \rangle\!_{I}\\
        &+ \frac{2}{(i)^2} {}_{I}\!\langle [[C^{(M)\dagger}_{j\sigma'} C^{(M)}_{i\sigma}, \dot{\hat{V}}_I(t)], \hat{V}_I(t)] \rangle\!_{I} + \frac{1}{(i)^2} {}_{I}\!\langle [[C^{(M)\dagger}_{j\sigma'} C^{(M)}_{i\sigma}, \hat{V}_I(t)], \dot{\hat{V}}_I(t)] \rangle\!_{I}.
    \end{aligned}
\end{equation}
The reason for Eq.~(\ref{eq: Third time derivative of trace}) being zero is that the diagonal elements of $\dddot{\rho}|_{t=0}$ are all zero, just as for $\ddot{\rho}|_{t=0}$. Hence, we can see that the entropy goes at least as $t^4$. However, numerically, we find that, for the initial SD state in the original basis $\ket{\text{SD}} = c^{\dagger}_{i_1\sigma_1}\cdots c^{\dagger}_{i_N\sigma_N} \ket{0}$, the entropy for short times evolves as 
\begin{equation}\label{eq: Entropy scaling Hubbard SD original basis}
    \begin{aligned}
        S^{(1)}_{2}(t) \approx \ln{N} + g_1\, \tau^4U^{2}t^6,
    \end{aligned}
\end{equation}
where $g_1$ is a constant independent of $\tau,U$ and $t$ (it can depend on $N,D$).

\section{Symmetry constraints on entanglement bounds}
\label{app:Symm}

The symmetry of a state considered above Eq.~(\ref{eq: Refined upper bound to entropy}) of the main text,  $\hat{N}_{\sigma}\ket{\Psi} = N_{\sigma}\ket{\Psi}$, gives us a block-diagonal form for $\rho^{(M)}$. We can understand this as a selection rule in the language of tensor operators for the Abelian symmetry group $U(1)^l$. For each conserved charge $\hat N_\sigma$ ($\sigma=1,\dots,l$), we call an operator $T_{\vec q}$ a tensor operator of rank $\vec q=(q_1,\dots,q_l)$ if $[\hat N_\sigma,\,T_{\vec q}] = q_\sigma\,T_{\vec q}$. The creation and annihilation operators carry definite ranks:  
\begin{equation}\label{eq: Commutator of creation and annihilation ops with charge}
    [\hat N_\sigma,\,c^\dagger_{a\sigma'}]=\delta_{\sigma\sigma'}\,c^\dagger_{a\sigma'} \,, \,\,\,\, [\hat N_\sigma,\,c_{a\sigma'}]=-\delta_{\sigma\sigma'}\,c_{a\sigma'}.
\end{equation}
Hence, any string of creation and annihilation operators of the form $O_{I;J}=c^\dagger_{a_1\sigma_1}\cdots c^\dagger_{a_M\sigma_M}\,c_{a'_M\sigma'_M}\cdots c_{a'_1\sigma'_1}$ gives 
\begin{equation}\label{eq: Commutator of the DM with the charge}
    [\hat N_\sigma,\,O_{I;J}]=(m_\sigma(I)-m_\sigma(J))O_{I;J},
\end{equation}
where $m_\sigma$ counts the number of indices with internal label $\sigma$ in the corresponding multi-index $I=a_1\sigma_1, \cdots, a_M\sigma_M$, $J = a'_1\sigma'_1, \cdots, a'_M\sigma'_M$. As a result, $O_{I;J}$ is a rank $\vec{q}$ tensor operator with components $q_\sigma=m_\sigma(I)-m_\sigma(J)$. 

In a fixed-charge eigenstate $\hat N_\sigma\ket{\Psi}=N_\sigma\ket{\Psi}$, we get a selection rule for matrix elements of $M$-body DM. Indeed, taking the expectation value of both sides of the Eq.~(\ref{eq: Commutator of the DM with the charge}) with respect to the state $\ket{\Psi}$, we see that the matrix elements vanish unless $\vec q=\vec 0$, implying $m_\sigma(I)=m_\sigma(J)$ for all $\sigma$. This is precisely the block-diagonal structure of $\rho^{(M)}$, with blocks labeled by the integers $(m_1,\dots,m_l)$ satisfying $\sum_{\sigma=1}^l m_\sigma=M$. 

We note that the same representation-theoretic viewpoint extends to the non-Abelian symmetries (e.g. $SU(2)$) by decomposing operator strings into higher rank irreducible tensor operators. The selection rules still constrain which components contribute. However, in that setting, multiplicity spaces and state-dependent irrep weights generally obstruct a closed-form combinatorial determination of block sizes and traces analogous to the $U(1)^l$ case.

Since the string of $m_{\sigma}$s uniquely identifies each block, we can use the notation $\hat{\rho}^{(M)}(m_1, \dots, m_l)$ for each generic block. The number of such blocks is thus the number of unique strings of $m_{\sigma}$s with the constraint $\sum_{\sigma=1}^l m_\sigma=M$ and is given by
\begin{equation}\label{eq: number of unique sets corresponding blocks}
    \begin{aligned}
        n_{\text{blocks}} = \binom{M+l-1}{l-1}.
    \end{aligned}
\end{equation}

Next, we determine the trace and size of a given block $\hat{\rho}^{(M)}(m_1, \dots, m_l)$ using a purely combinatorial approach. 

\begin{figure}[t]
    \centering
    \subfigure[]{
        \includegraphics[width=0.48\columnwidth]{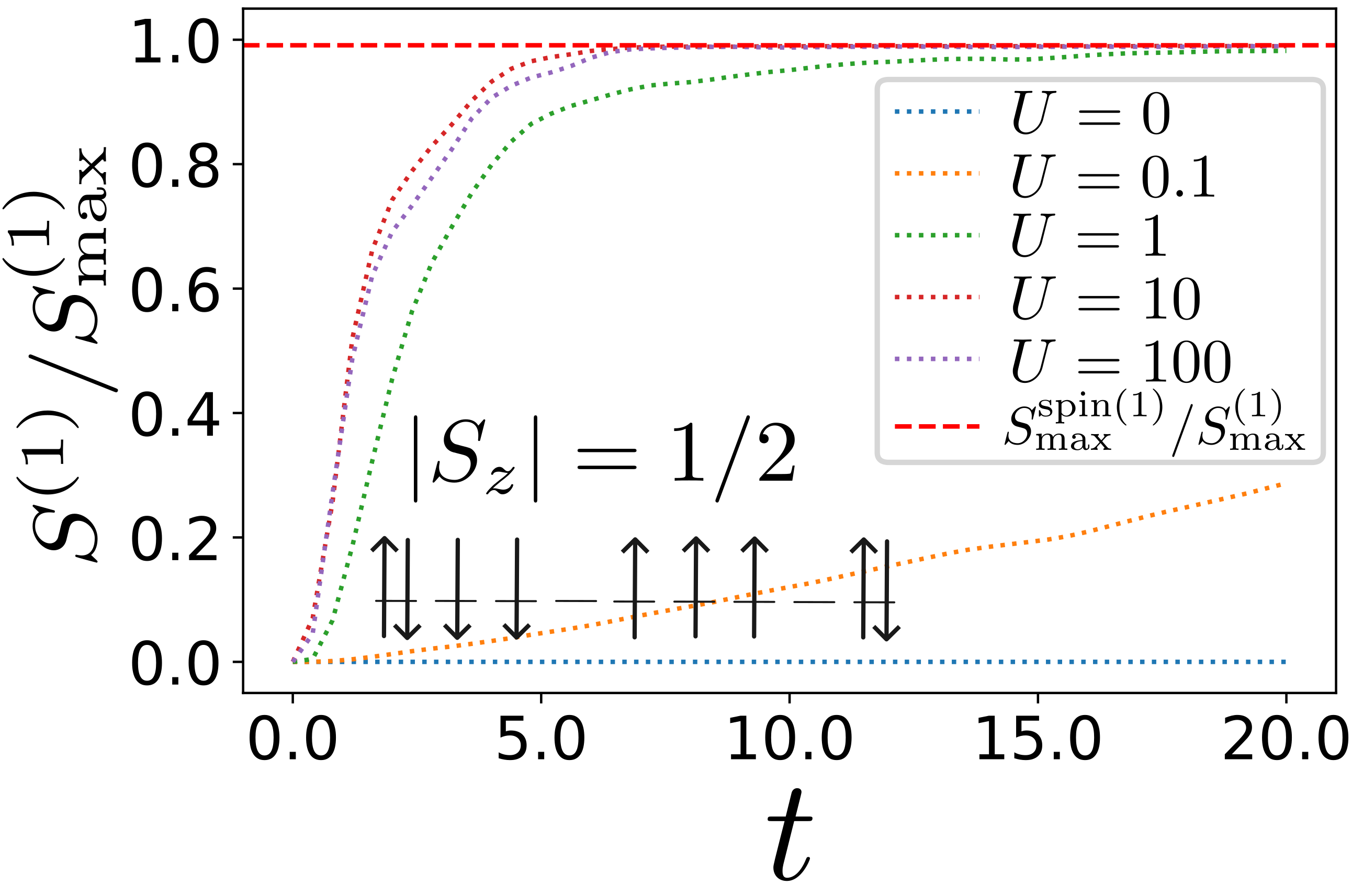}
        \label{fig:SD_4861}
    }
    \hspace{-1em}
    \subfigure[]{
        \includegraphics[width=0.48\columnwidth]{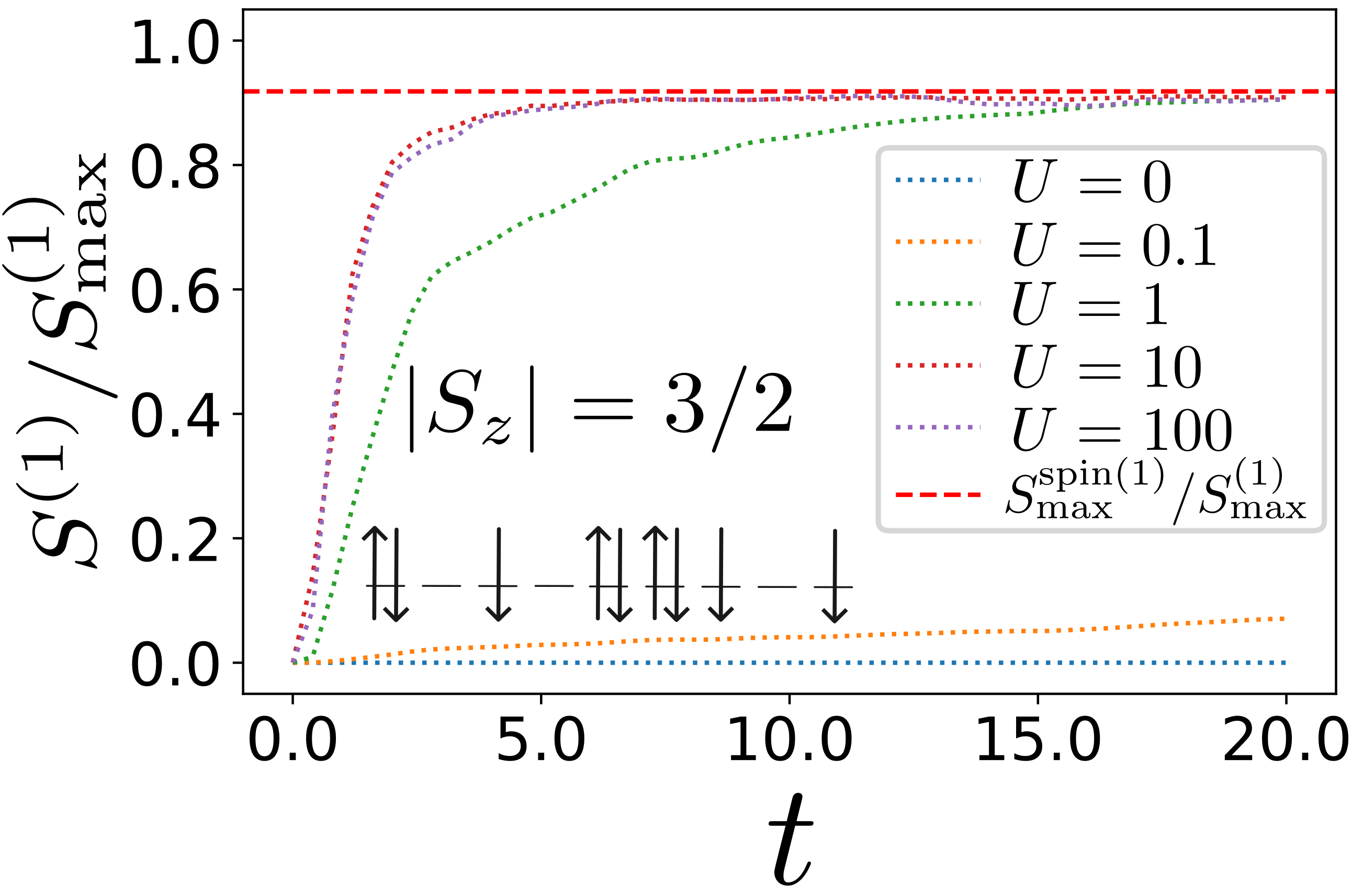}
        \label{fig:SD_14581}
    }
    \\[-1em] 
    \subfigure[]{
        \includegraphics[width=0.48\columnwidth]{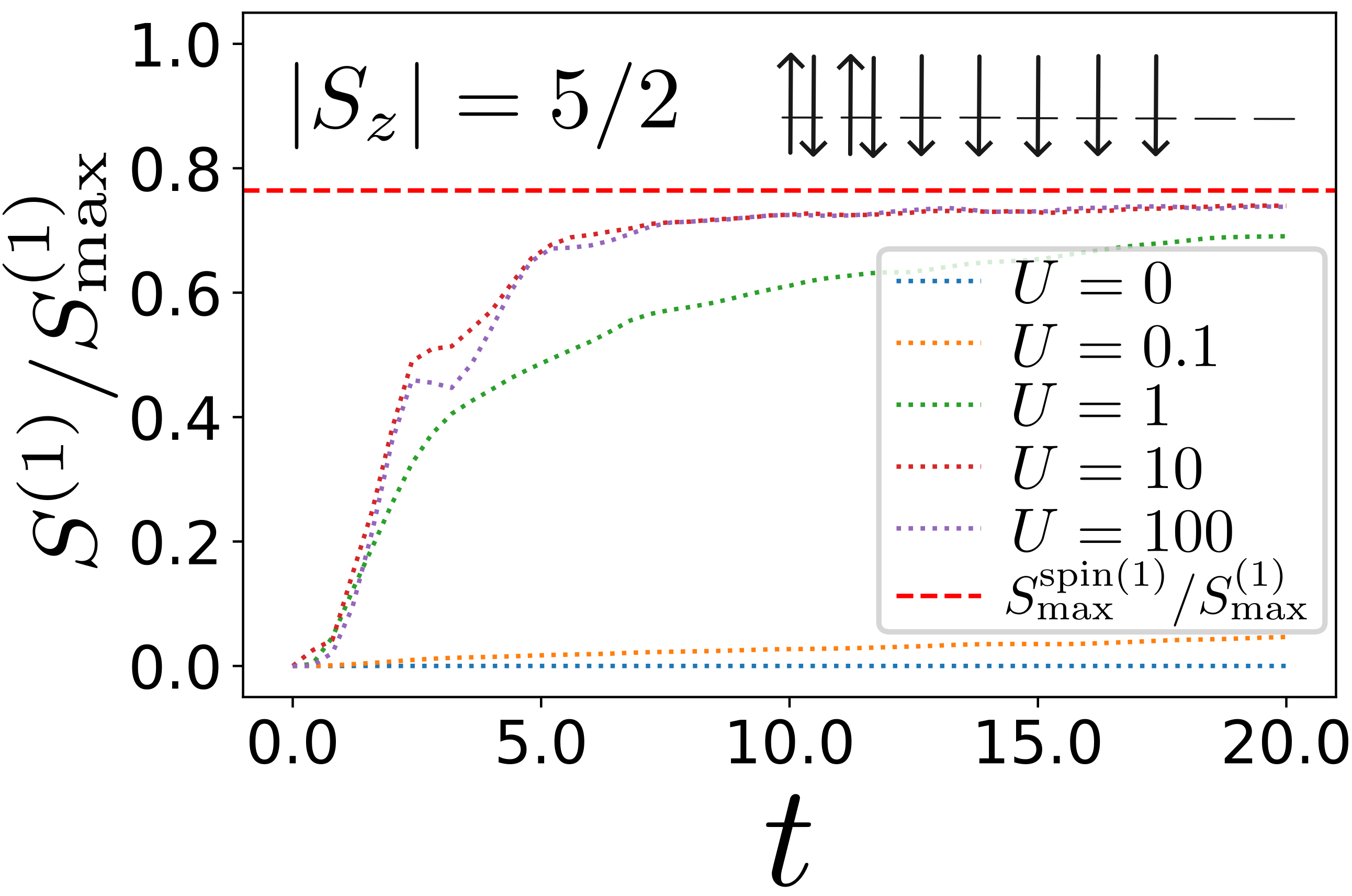}
        \label{fig:SD_19}
    }
    \hspace{-1em}
    \subfigure[]{
        \includegraphics[width=0.48\columnwidth]{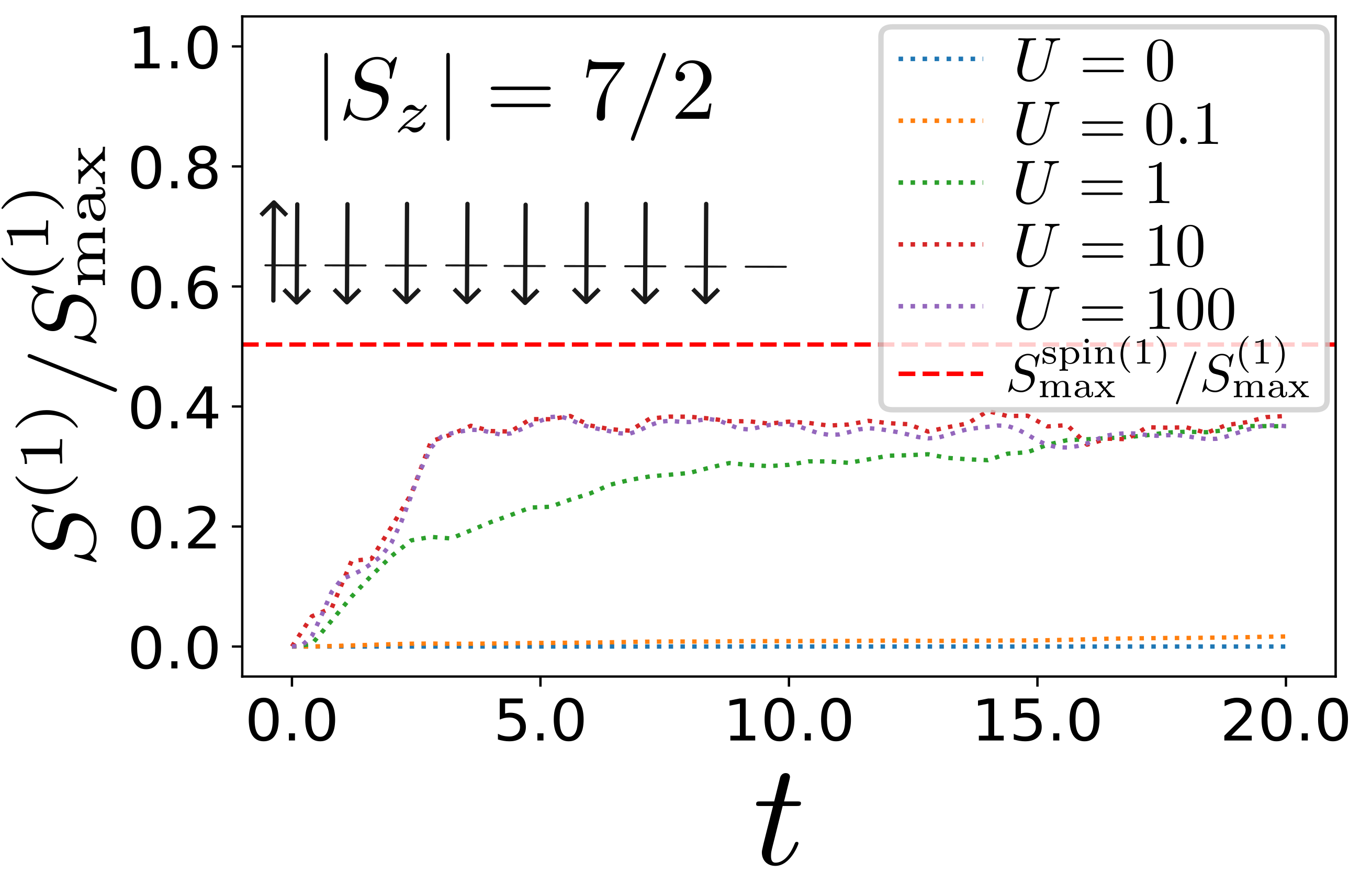}
        \label{fig:SD_1}
    }
    \caption{Hubbard dynamics and entropy saturation. Time evolution of the one-body entanglement entropy for a 9-site 1D lattice at half-filling, starting from SD initial states in the spin-projection sectors $|S_z|$ of 1/2, 3/2, 5/2, and 7/2 (in units of $\hbar$). Symmetry refined upper bounds to the entanglement entropy of Eq.~(\ref{eq: Refined upper bound to entropy})
    are given by red dashed lines.  
    }
    \label{fig: Full time evolution Hubbard only}
\end{figure}

\begin{figure*}[t]
    \centering
    \subfigure[]{    
        \centering
        \includegraphics[width=0.4\columnwidth]{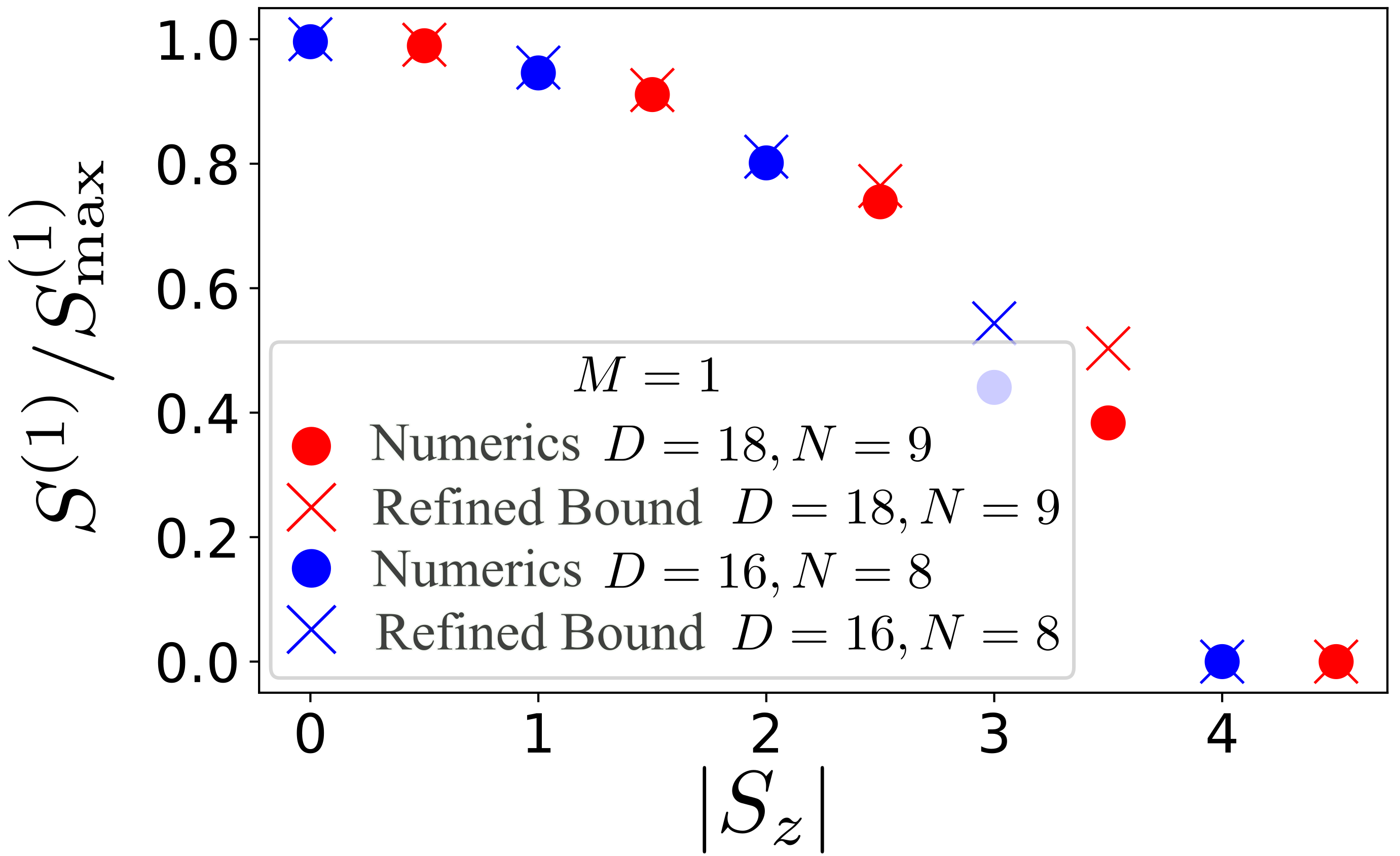}
        \label{fig:1-body entropy vs spin}
    }
    \hspace{0 \columnwidth}
    \subfigure[]{
        \includegraphics[width=0.4\columnwidth]{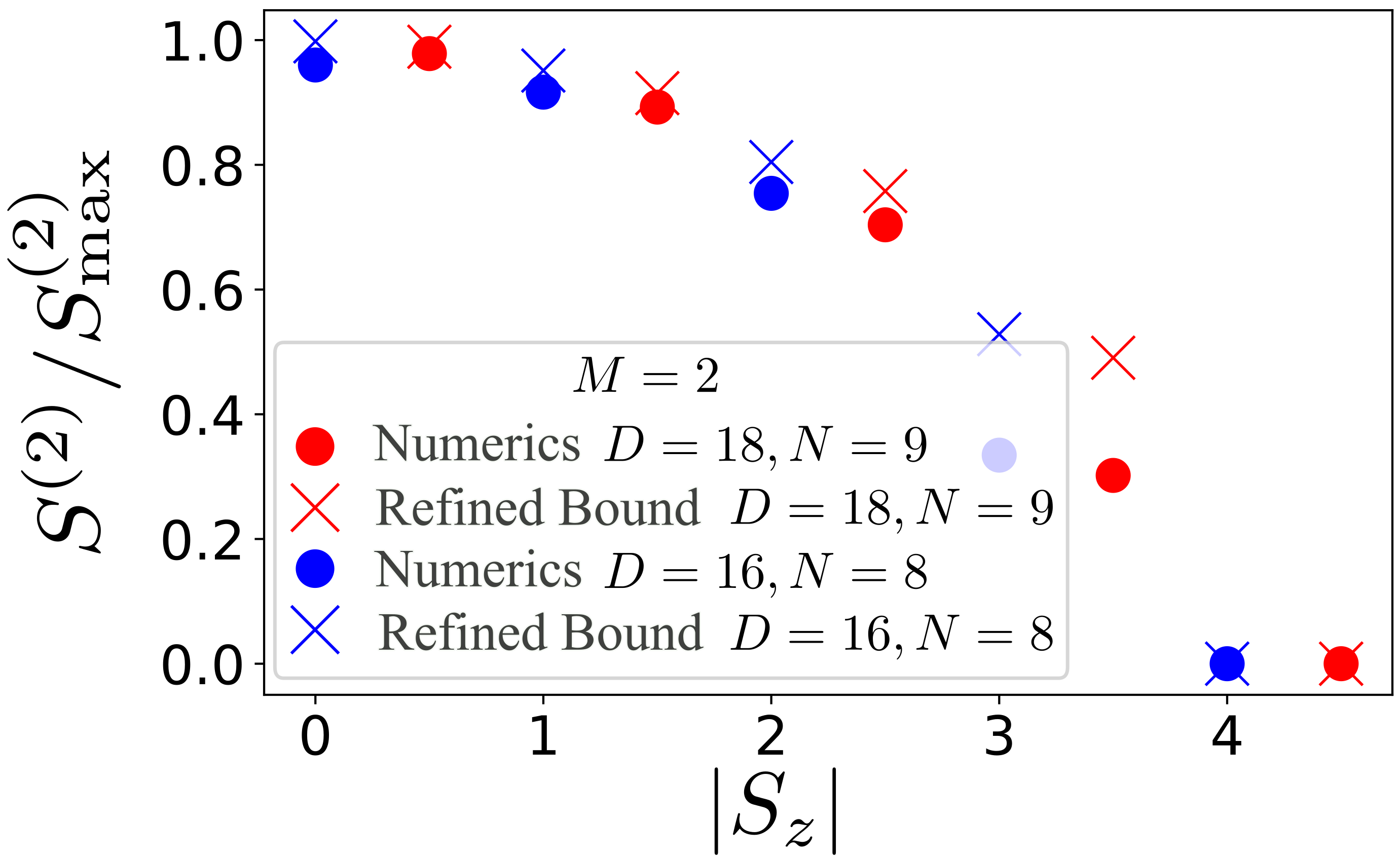}
        \label{fig:2-body entropy vs spin}
    }
    \hspace{0 \columnwidth}
    \subfigure[]{
        \includegraphics[width=0.4\columnwidth]{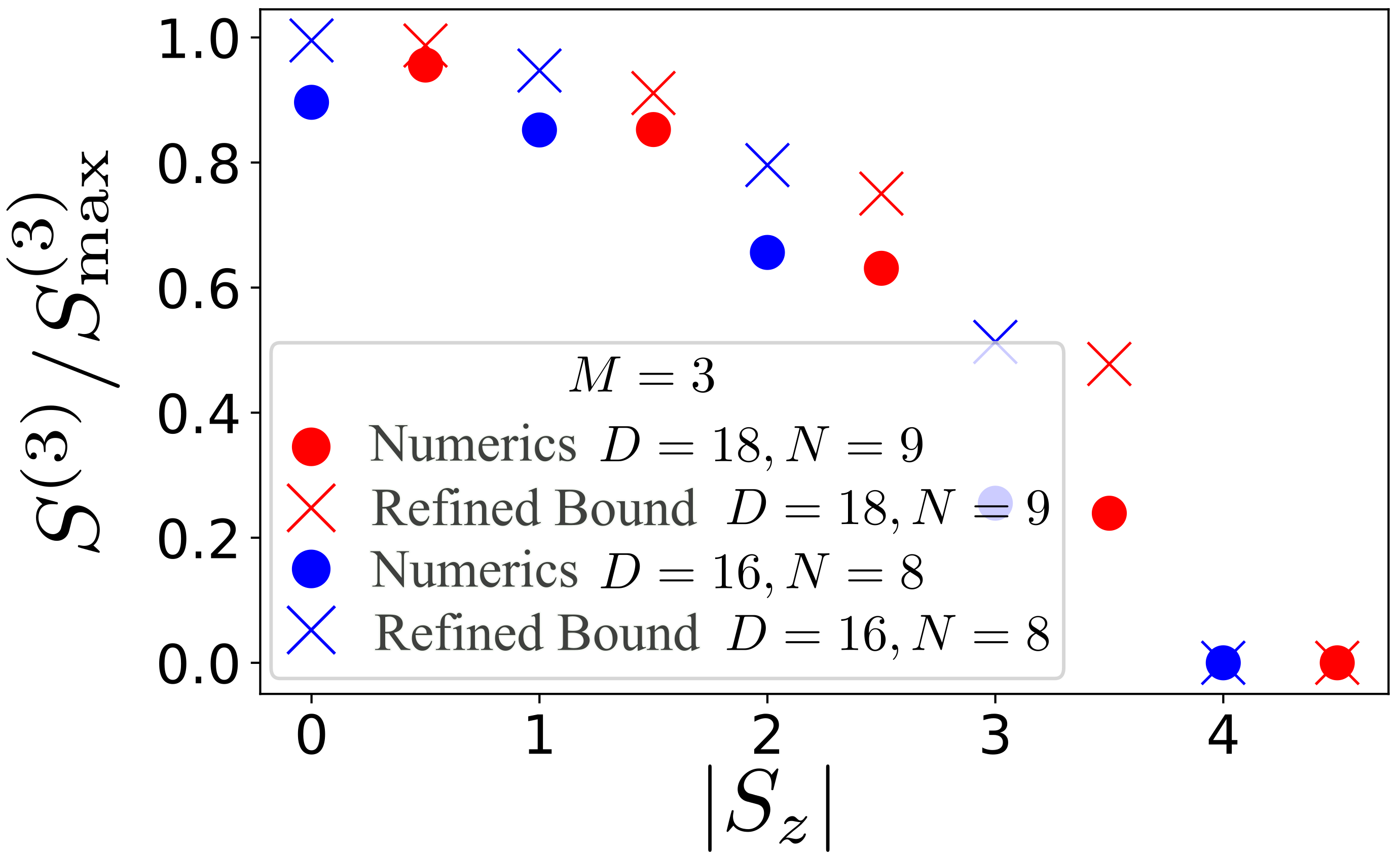}
        \label{fig:3-body entropy vs spin}
    }
    \caption{(a)-(c) Dependence of $M$-body entanglement entropy upper bound on  $|S_z|$ (in units of $\hbar$) of the initial SD state in 1D Hubbard model, given as a fraction of the absolute upper bound. Two distinct $D,N$ values are color coded. The circles correspond to numerically obtained entropy saturation values, similar to Fig.~\ref{fig: combined plots refined upper bound}, while crosses are theoretical refined bounds obtained from Eq.~(\ref{eq: Refined upper bound to entropy}). 
    }
    \label{fig:Entropy vs spin}
\end{figure*}

\subsection{Trace and size of a block}

\subsubsection{Trace of a block} First, let us consider the state to be $N$-fermion Slater determinant (SD) with $N_{\sigma}$ fermions having internal label $\sigma$ for all $\sigma \in \{1, \dots, l\}$ such that $N = \sum\limits_{\sigma = 1}^{l} N_{\sigma}$. Then the contribution to the block $\hat{\rho}^{(M)}(m_1 \dots m_l)$ is $\prod\limits_{\sigma=1}^{l}\binom{N_{\sigma}}{m_{\sigma}}$ unit entries on the diagonal. This is the case because $m_{\sigma}$ fermions with internal label $\sigma$ can be chosen $\binom{N_{\sigma}}{m_{\sigma}}$ ways out of $N_{\sigma}$ internal label-$\sigma$ fermions of SD state. (Here, we only consider the case where $m_{\sigma}\leq N_{\sigma}$ for all $\sigma \in \{1, \dots, l\}$, otherwise every element of $\hat{\rho}^{(M)}(m_1 \dots m_l)$ would be zero and will not contribute to the entropy calculations.) Each realization gives 1 on the diagonal. Hence, for the trace we get
\begin{equation}\label{eq: Trace of the block general appendix}
    \begin{aligned}
        \text{Tr}\hat{\rho}^{(M)}(m_1 \dots m_l) = \prod\limits_{\sigma=1}^{l}\binom{N_{\sigma}}{m_{\sigma}}.
    \end{aligned}
\end{equation}
Now let us say that we have a generic state $\ket{\Psi} = \sum_k \alpha_k \ket{\text{SD}_k}$ which has a definite $\sigma$-sector for all $\sigma$, with each $\ket{\text{SD}_k}$ having the same number of $N_1, \dots ,N_{\sigma}, \dots , N_l$. The contribution to diagonal elements of the block, coming from $k^{\text{th}}$ SD, is $\prod\limits_{\sigma=1}^{l}\binom{N_{\sigma}}{m_{\sigma}}$ many $|\alpha_k|^2$ for each $k$. Hence, from the normalization of $\ket{\Psi}$, the trace is still given by Eq.~(\ref{eq: Trace of the block general appendix}).

\subsubsection{Size of a block} In the case of $D/l$ sites on the lattice, since all sites can accommodate fermions with any $\sigma$ internal label, we have $\binom{D/l}{m_{\sigma}}$ ways to arrange the $m_{\sigma}$ fermions with internal label $\sigma$ on $D/l$ lattice sites $\forall \sigma$. Hence, the size of the block is
\begin{equation}
    \begin{aligned}
        n_{\text{size}}(m_1 \dots m_l) = \prod\limits_{\sigma=1}^{l} \binom{D/l}{m_{\sigma}}.\label{eq: Size of the block general appendix}
    \end{aligned}
\end{equation}
It is important to note that $m_{\sigma} \leq D/l, \,\,\,\,\, \forall\sigma$. This follows from the already established condition $m_{\sigma}\leq N_{\sigma}$ and the fact that $N_{\sigma} \leq D/l$ for all $\sigma \in \{1, \dots, l\}$. Indeed, if $N_{\sigma} > D/l$, then there would exist at least one site with more than one fermion with internal label $\sigma$, forbidden by Pauli exclusion. 

\subsection{Maximimum entropy in the presence of symmetries}

We now proceed to the refined upper bound of the $M$-body entanglement entropy, which in the presence of symmetries decomposes into a sum over contributions of distinct blocks. The maximum von Neumann entropy for an $n \times n$ DM $\rho$, is given by $S_{\text{max}} = \text{Tr}\rho \ln (\frac{n}{\text{Tr}\rho})$. Then we can write the maximum entropy for a given block $\hat{\rho}^{(M)}(m_1 \dots m_l)$
\begin{equation}\label{eq: Max entropy of the block general}
    \begin{aligned}
        S^{(M)}_{\text{max}}(m_1 \dots m_l) = \left[\prod\limits_{\sigma=1}^{l}\binom{N_{\sigma}}{m_{\sigma}}\right] \ln \left(\frac{\prod\limits_{\sigma=1}^{l}\binom{D/l}{m_{\sigma}}}{\prod\limits_{\sigma=1}^{l}\binom{N_{\sigma}}{m_{\sigma}}}\right)
    \end{aligned}
\end{equation}
The symmetry refined upper bound to the entropy then becomes 
\begin{equation}
    \begin{aligned} \label{eq:SMaxSymm}
        S^{(M)}_{\text{max}}(N_1, \dots, N_l) = \sum_{m_1, \dots, m_l}'\left[\prod\limits_{\sigma=1}^{l}\binom{N_{\sigma}}{m_{\sigma}}\right] \ln \left(\frac{\prod\limits_{\sigma=1}^{l}\binom{D/l}{m_{\sigma}}}{\prod\limits_{\sigma=1}^{l}\binom{N_{\sigma}}{m_{\sigma}}}\right),
    \end{aligned}
\end{equation}
where the prime indicates that the sum is constrained by $\sum\limits_{\sigma=1}^{l}m_{\sigma} = M$ and $N_{\sigma} \geq m_{\sigma}$, $\forall \sigma \in \{1, \dots, l\}$.

Taking $l=2$ in Eq.~(\ref{eq:SMaxSymm}) concludes the derivation of Eq.~\eqref{eq: Refined upper bound to entropy} of the main text. Fig.~\ref{fig: Full time evolution Hubbard only} shows the time evolution of one-body entanglement for several different initial Slater determinant states under Hubbard dynamics and how the upper bound is refined. Fig.~\ref{fig:Entropy vs spin} includes other $D,N,M$ values and shows corresponding refined upper bounds.

\end{widetext}

\end{document}